%% file: main_arxiv.tex
\documentclass[aps,pre,twocolumn,nofootinbib,10pt,superscriptaddress]{revtex4-1}

\usepackage{graphicx}
\usepackage{dcolumn}
\usepackage{bm}

\usepackage[utf8]{inputenc}
\usepackage{mathtools}
\usepackage[T1]{fontenc}
\usepackage{mathptmx}
\usepackage[utf8]{inputenc}
\usepackage{graphicx}
\usepackage{subcaption}
\usepackage{amsmath}
\usepackage{amssymb}
\usepackage{dcolumn}
\usepackage{bm}
\usepackage{xr}
\usepackage{cleveref}
\usepackage{soul}
\usepackage{color}
\usepackage{ulem}
\usepackage[usenames,dvipsnames]{xcolor}

\begin{document}

\title{LCST and Closed-Loop Phase Behavior in Non-Associating Fully Symmetric Multicomponent Polymer Systems}
\author{Artem Petrov}
 \email{aipetrov@mit.edu}
	\affiliation{Department of Chemical Engineering, Massachusetts Institute of Technology, Cambridge, Massachusetts 02139, United States}
 \author{Alfredo Alexander-Katz}
\email{aalexand@mit.edu}
	\affiliation{Department of Materials Science and Engineering, Massachusetts Institute of Technology, Cambridge, Massachusetts 02139, United States}
	\date{\today}
	
\begin{abstract}

Certain low-molecular liquids and multicomponent polymer systems can phase separate upon heating, exhibiting a lower critical solution temperature (LCST). Moreover, a narrow class of materials can undergo disordering transition upon further heating, yielding closed-loop phase diagrams. Previously, it was shown that LCST or closed-loop phase behavior can appear in the models of liquids in which components are asymmetric or interaction potentials have a specifically designed attraction. Here, we show theoretically that LCST and closed-loop phase behavior can occur in a significantly wider set of models. In particular, we found that these phenomena can be exhibited by the simplest and widely used coarse-grained (CG) models of any multicomponent liquid in which species are \textit{fully symmetric} and where all particles interact via an \textit{arbitrary repulsive} potential. In addition, we discovered that LCST and closed-loop phase behavior in these models emerges merely due to the basic property of CG liquid models, namely, the appearance of strong monomer-monomer positional correlations at low monomer number density $\rho$. To demonstrate this, we simulated the models of fully symmetric binary blends and diblock copolymer melts where nonbonded monomers interacted via a generic $T$-independent purely repulsive harmonic potential. As predicted, LCST and closed-loop phase behavior emerged at $\rho$ sufficiently low to cause strong monomer-monomer correlations leading to a strong $T$-dependence of the effective coordination number, which, in turn, induced a nonmonotonic $T$-dependence of the Flory-Huggins parameter. To summarize, we discovered that the simplest fully symmetric non-associating CG models of multicomponent liquids can exhibit complex temperature response (LCST and closed-loop phase behavior) due to a mechanism stemming from the basic nature of any CG liquid model. Due to the simplicity of the models, this mechanism might contribute to the emergence of LCST and closed-loop phase behavior in many existing polymer materials.

\end{abstract}

\maketitle

\section{Introduction}

Typically, macro- and micro-phase separated multicomponent materials (polymer blends, low-molecular and polymer solutions, block copolymer melts, etc.) transition into homogeneous state upon heating. As a result, the upper critical solution temperature (UCST) appears in the phase diagrams of such systems \cite{lee2009phase}. However, there exists a class of liquids where the trend is reversed: an increase of temperature causes phase separation, which leads to the emergence of a lower critical solution temperature (LCST). Such behavior was observed for water-triethylamine solutions \cite{erikson2017partially}, for solutions of polystyrene (PS) in ethyl acetate \cite{bae1993representation}, for blends of PS and poly(vinyl methyl ether) \cite{rubinstein2003polymer}, and for PS-block-poly(n-butylmethacrylate) block copolymer melts \cite{russell1994lower,yeung1994lower}. Moreover, there exists a subclass of LCST systems that have a finite temperature range in which the phase separated state exists (i.e., a UCST appears above an LCST). Phase diagrams of such liquids exhibit "closed-loop" regions of phase separation. Such systems include water-nicotine solutions \cite{bailey2018dynamics}, aqueous solutions of poly(ethylene oxide) \cite{knychala201750th}, blends of deuterated PS and poly(n-pentylmethacrylate) (PnPMA) \cite{ryu2004complex}, and PS-block-PnPMA block copolymer melts \cite{yeol2002closed}. Polymer materials having an LCST and/or closed-loop phase behavior are widely used as tunable "smart" thermoresponsive materials, especially in biomedicine \cite{russell1994lower,yuan2023thermoresponsive}.

The physical explanation of the emergence of LCST and closed-loop phase behavior is, probably, one of the oldest problems in polymer science and liquid state theory. One of the first discovered mechanisms attributed such behavior to the existence of association between the species \cite{goldstein1983theory,yuan2023thermoresponsive,oh1998liquid}. For instance, in the classical example of water-poly(N-isopropylacrylamide) solution, heating disrupts hydrogen bonds between monomers and water, which leads to macro-phase separation and appearance of an LCST \cite{yuan2023thermoresponsive}. However, specific association cannot explain the emergence of LCST in nonpolar systems such as blends of polyisobutylene and deuterated polypropylene \cite{rubinstein2003polymer}. As a result, numerous theories were constructed that provided alternative explanations of LCST and closed-loop phase behavior in polymer systems and low-molecular liquids. The celebrated Sanchez-Lacombe theory \cite{sanchez1976elementary,sanchez1978statistical}, free volume approach \cite{patterson1969free}, Ruzette and Mayes theory \cite{ruzette2001simple}, and Born-Green-Yvon-type lattice theories \cite{lipson1992born,white2012pure,clark2012lcst} attributed the emergence of LCST to the finite compressibility of the liquids. They proved that in compressible systems, the difference between the equation of state properties of pure components (we will further refer to this difference as "asymmetry" between species) leads to the emergence of an LCST in any liquid \cite{sanchez1978statistical}. Moreover, the lattice cluster theory of Freed and coworkers \cite{dudowicz1993relation} showed that the existence of LCST and closed-loop phase behavior is possible even in incompressible polymer blends and diblock copolymer (diBCP) melts. Instead, the appearance of these phenomena was attributed to the structural asymmetry between monomers of different species \cite{dudowicz2002new}. Therefore, previous theoretical approaches typically explained LCST and closed-loop phase behavior either via the existence of special association between components or to a certain asymmetry between the species; on the other hand, non-associating "fully symmetric" systems, which state does not change upon relabeling the species due to identical structure and volume fractions of the components, were generally predicted to exhibit UCST phase behavior.

Besides investigating the non-UCST phase behavior analytically, researchers studied it in computer simulations. When certain liquids were modeled realistically with near-atomistic accuracy, LCST and closed-loop phase behavior could be observed in a simulation \cite{shahamat2013molecular,oh2012role}. Unfortunately, due to the complexity of the models, the downside of such detailed simulation is the lack of clarity about a particular microscopic mechanism causing the non-UCST phase behavior. As a result, other researchers performed coarse-grained (CG) modeling, in which they observed non-UCST phase behavior either due to realistically \cite{fayaz2022coarse,perez2020coil} or coarsely \cite{azizi2022reentrant,gromov1998simulation,luna1997polymer} modeled structural difference between the species or due to the introduction of temperature-dependent potentials \cite{jiang2025temperature,van2000pressure,abbott2015temperature}.
Moreover, researchers observed LCST and closed-loop phase behavior in the models having specifically designed association between unlike particles \cite{davies1999simulation,davies2000closed,kotelyanskii1998phase,belousov2008global,thamm2003phase,schweizer1993analytic,lopes1999phase,materniak2013reentrant,almarza2001reentrant,toda2016phase}. In the majority of these works, the complex non-UCST phase behavior occurred due to directional bonds (with temperature-dependent association constants) \cite{davies1999simulation,davies2000closed,kotelyanskii1998phase,belousov2008global,thamm2003phase}. Also, there exist a handful of studies demonstrating closed-loop phase behavior in the models having spherically symmetric interaction potentials with peculiarly constructed attraction between unlike particles \cite{materniak2013reentrant,almarza2001reentrant,schweizer1993analytic,lopes1999phase}. In these works, LCST appeared due to 
an explicitly introduced special mismatch between attraction strength, repulsion range, and attraction range of the interaction potentials between like and unlike particles. Similar to directional/hydrogen bonds, this fine-tuned choice of potentials favored phase separation at high $T$ and association between unlike beads and the formation of a (non-randomly mixed) homogeneous phase at low $T$ \cite{materniak2013reentrant,almarza2001reentrant,schweizer1993analytic,lopes1999phase}.
Thus, the previous modeling effort qualitatively confirmed the theoretical picture in which LCST and closed-loop phase behavior emerge in the systems that have either asymmetry between species, specifically designed association between unlike particles, or temperature-dependent interaction.

In this work, we show that LCST and closed-loop phase behavior can appear in a significantly broader set of systems. We studied a class of simple coarse-grained (CG) multicomponent polymer liquid models that did not possess model features outlined in the previous paragraph and, therefore, were intuitively expected to have a usual UCST phase behavior. In particular, polymer chains were represented by a beads-on-a-spring model, monomers and chains of all species were identical, and nonbonded monomers interacted via an arbitrary purely repulsive temperature-independent potential. A CG "monomer" in these models represented a group of real monomers ($\gtrsim 1$ Kuhn segment). It is worth mentioning that these models are widely utilized to simulate multicomponent polymer and low-molecular liquids \cite{petrov2024simple,petrov2026universality,medapuram2015universal,glaser2014collective,glaser2014universality,groot1997dissipative,minkara2019new} for example, in dissipative particle dynamics (DPD) \cite{groot1997dissipative,minkara2019new}. We analytically derived the temperature dependence of the universal effective Flory-Huggins parameter $\chi_e$ that governs the phase behavior in these models. The shape of this $\chi_e(T)$ function was predicted to be mainly controlled by the $T$-dependence of the effective coordination number $z_\infty$. We showed that a superlinear $z_\infty(T)$ dependency in these models was induced just by the onset of strong positional correlations between interacting monomers at low monomer number density $\rho$. In turn, this $z_\infty(T)$ behavior was predicted to lead to LCST and closed-loop phase behavior regardless the type of a multicomponent polymer system, the form of the repulsive nonbonded interaction, and the compositional symmetry. To confirm these predictions, we performed computer simulations of low-$\rho$ fully symmetric homopolymer blend and diBCP melt models with purely repulsive harmonic nonbonded potential and discovered their LCST and closed-loop phase behavior. At high $\rho$, on the other hand, such models exhibited UCST phase behavior, which was also predicted by the theory. As a result, we discovered an effect that can produce LCST and closed-loop phase behavior in a wide class of non-associating CG models of multicomponent polymer liquids, including the simplest fully symmetric models in which this behavior was not expected from the previously developed theories. Due to its general and basic nature, this effect might contribute to the temperature response of many experimentally studied systems.

\section{Theory of Temperature-Dependent Phase Behavior in Coarse-Grained Models}


We focused on CG models where monomers were represented as point-like particles in continuum space interacting via a purely repulsive nonbonded potential $U_\text{nonbonded}=a_{ij}v(r_{ij})$, where $a_{ij}$ are the energy coefficients depending on the type of interacting monomers $i$ and $j$, and $v(r_{ij})$ is a dimensionless monotonically decreasing function of distance between the monomers. The number of monomers per unit volume $\rho$ was a fixed constant. In this work, we reviewed only structurally symmetric models, in which the same bond potential was applied to all species, $a_{AA}=a_{BB}\equiv a_{xx}$, and all chains (for example, homopolymers or block copolymers) had the same degree of polymerization $N$. For any studied model, we fixed the $a_{AB}$ coefficient such that $a_{AB}>a_{xx}$ to potentially allow segregation between $A$ and $B$ species. 

We developed a theory for the temperature response of such CG models assuming that the compressibility effects play minor role in determining a model's phase behavior. This can be achieved, for instance, by making the models compositionally symmetric, which was done in the simulations described in the next section; however, nearly incompressible compositionally asymmetric models are also described by the theory. In turn, in the structurally symmetric, nearly incompressible binary polymer melts, the thermodynamic state of the system is assumed to be universal, i.e., it is controlled by only three effective parameters: chain architecture, $\chi_eN$, and the invariant chain length $\bar{N}$, which is proportional to the average squared number of chains situated in a single-chain volume in a melt \cite{petrov2026universality,petrov2024simple,medapuram2015universal,glaser2014universality}. Recently, we analytically derived the definition of $\chi_e$ as a function of CG model parameters that produces such universal thermodynamic behavior in the CG models described above \cite{petrov2026universality}. According to this definition, $\chi_e$ is an alternating sum in powers of the mismatch energy $\alpha=a_{AB}-a_{xx}>0$ (Eq. \ref{eq1}).

\begin{equation}
\label{eq1}
    \chi_e(\alpha) = \sum_{n=1}^{\infty}\frac{(-1)^{n+1}}{n!}\left(\frac{\alpha}{k_B T}\right)^n\kappa_n\left[z\right]_{0,N\to\infty}
\end{equation}

In Eq. \ref{eq1}, $\kappa_n\left[z\right]_{0,N\to\infty}$ are the cumulants of the distribution of the effective coordination number $z$ defined as a sum over $v(r_{ij})$ values measured between a selected monomer and all other monomers located on different chains (Eq. \ref{eq2}). 

\begin{equation}
\label{eq2}
    z_k\equiv\sum_{i:\text{chain}(i)\neq \text{chain}(k)} v(r_{ik})
\end{equation}

To calculate $\kappa_n\left[z\right]_{0,N\to\infty}$ in Eq. \ref{eq1}, one first needs to calculate $\kappa_n[z(N)]_0$: the cumulants of the distribution of $z$ (Eq. \ref{eq2}) obtained in the homogeneous state of the model (i.e., at $\alpha=0$) in which chains have the degree of polymerization $N$. After that, one has to extrapolate $\kappa_n[z(N)]_0$ to the infinite chain length by using Eq. \ref{eq3}.

\begin{equation}
\label{eq3}
\begin{cases}
    \langle z(N) \rangle_0 =z_\infty\left(1+\frac{(6/\pi)^{3/2}}{\rho b^3N^{1/2}}+\frac{\delta}{N}\right)\\
    \kappa_n[z(N)]_0 =\kappa_n[z]_{0,N\to\infty}\left(1+\frac{C}{\rho b^3N^{1/2}}\right),\quad n\geq 2
\end{cases}
\end{equation}

In Eq. \ref{eq3}, $C$ and $\delta$ are model-dependent fitting constants, and $b$ is the statistical segment length determined by extrapolating the scaled average squared radius of gyration of a chain ($6R_g^2/N$) measured at $\alpha=0$ to $N\to\infty$ according to Eq. \ref{eq4}, where $\gamma$ is a model-dependent fitting constant \cite{glaser2014universality,glaser2014collective,qin2009renormalized}:

\begin{equation}
\label{eq4}
    \frac{6R^2_g(N)}{N}=b^2\left(1-\frac{1.42}{\rho b^3N^{1/2}}+\frac{\gamma}{N}\right)
\end{equation}

We note in passing that $\bar{N}$ can be defined as $\bar{N}=N\rho^2b^6$; for simplicity, we will only review the models having weak $\bar{N}(T)$ dependency.

In what follows, we built a theory of the $T$-dependence of $\chi_e$ defined in Eq. \ref{eq1}. Despite the fact that the universality of multicomponent polymer systems (and, therefore, the definition of $\chi_e$ in Eq. \ref{eq1}) remains a hypothesis, we will use this definition as a convenient starting point to analyze the $T$-response of the CG models; the simulations of the studied models will verify the theoretical predictions (see next section). Moreover, all models studied in this work had small enough $\alpha$ (i.e., $\alpha<a_{xx}$), which guaranteed the convergence of Eq. \ref{eq1} and, therefore, its applicability to the investigated models \cite{petrov2026universality}.

Since, according to the universality hypothesis, $\chi_eN$ fully determines the thermodynamic state of the studied CG models (at a fixed chain architecture, composition, and $\bar{N}$), the $\chi_e(T)$ function completely characterizes the type (UCST, LCST, etc.) of temperature-driven phase transitions in such models. In turn, $\chi_e(T)$ is fully defined by the $T$-dependence of the distribution of $z$ in the "reference" state (at $\alpha=0$ and $N\to\infty$) of a given two-component melt model. For simplicity, we will further assume truncated nonbonded potential (i.e., $v(r)=0$ for $r\geq r_{cut}$). As a result, $z$ for a selected monomer in the "reference" state will be determined by two random variables: (i) the number $k_{neigh}$ of other-chain neighbors, i.e., the monomers situated within the cutoff distance $r_{cut}$ and on different chains from the selected monomer, and (ii) $v(r)$, where $r$ is measured between the selected monomer and any other-chain neighbor. Therefore, we focused on the estimation of the $T$-dependence of the "reference"-state distributions of $v$ and $k_{neigh}$ to determine the desired $\chi_e(T)$ dependence.

For a selected monomer and its other-chain neighbor, the probability density of observing distance $r$ between them is proportional to the exponent of the two-particle inter-chain potential of mean force $w(r)$: $P(r)\propto r^2 \exp[-w(r)/(k_BT)]$. In turn, $w(r)$ strongly depends on monomer density $\rho$. If $\rho$ is large, $w(r)$ can exhibit complex behavior depending on the chosen nonbonded potential \cite{likos2001criterion}. However, there exist simple soft repulsive potentials that produce a liquid with no positional correlations between monomers at infinitely high densities; for instance, the harmonic repulsive potential used in DPD \cite{groot1997dissipative} belongs to this class \cite{likos2001criterion}. For simplicity, we will consider only these potentials for the analysis of high-$\rho$ models. For such potentials, $w(r)\approx 0$ at high $\rho$, and $P(r)\propto r^2$ becomes roughly $T$-independent. Therefore, as $\rho\to\infty$, the distributions of $v(r)$ and $k_{neigh}$ (and, as a result, the $\kappa_n\left[z\right]_{0,N\to\infty}$ coefficients) become $T$-independent. Thus, the $T$-dependence of $\chi_e$ (Eq. \ref{eq1}) takes the following form (Eq. \ref{eq5}), where $C_n$ are model- and $\alpha$-dependent constants.

\begin{equation}
\label{eq5}
    \chi_e(T) \approx \sum_{n=1}^{\infty}\frac{(-1)^{n+1}}{n!}\frac{C_n}{T^n}, \quad \rho\to\infty
\end{equation}

\begin{figure}[!t]
\centering
  \begin{subfigure}{0.49\textwidth}\includegraphics[width=\linewidth,height=\textheight,keepaspectratio]{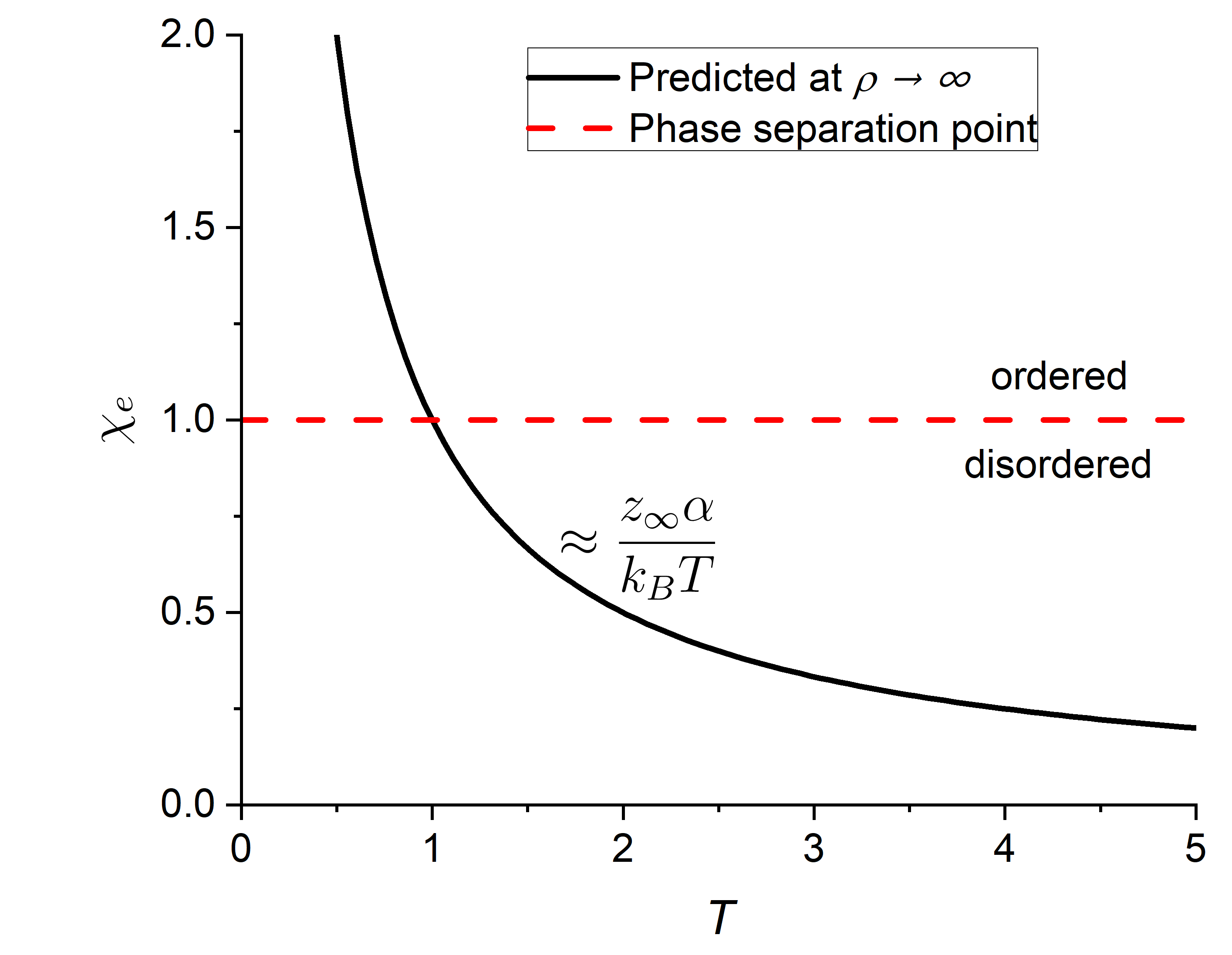}
	\caption{}
	\end{subfigure}
	\begin{subfigure}{0.49\textwidth}\includegraphics[width=\linewidth,height=\textheight,keepaspectratio]{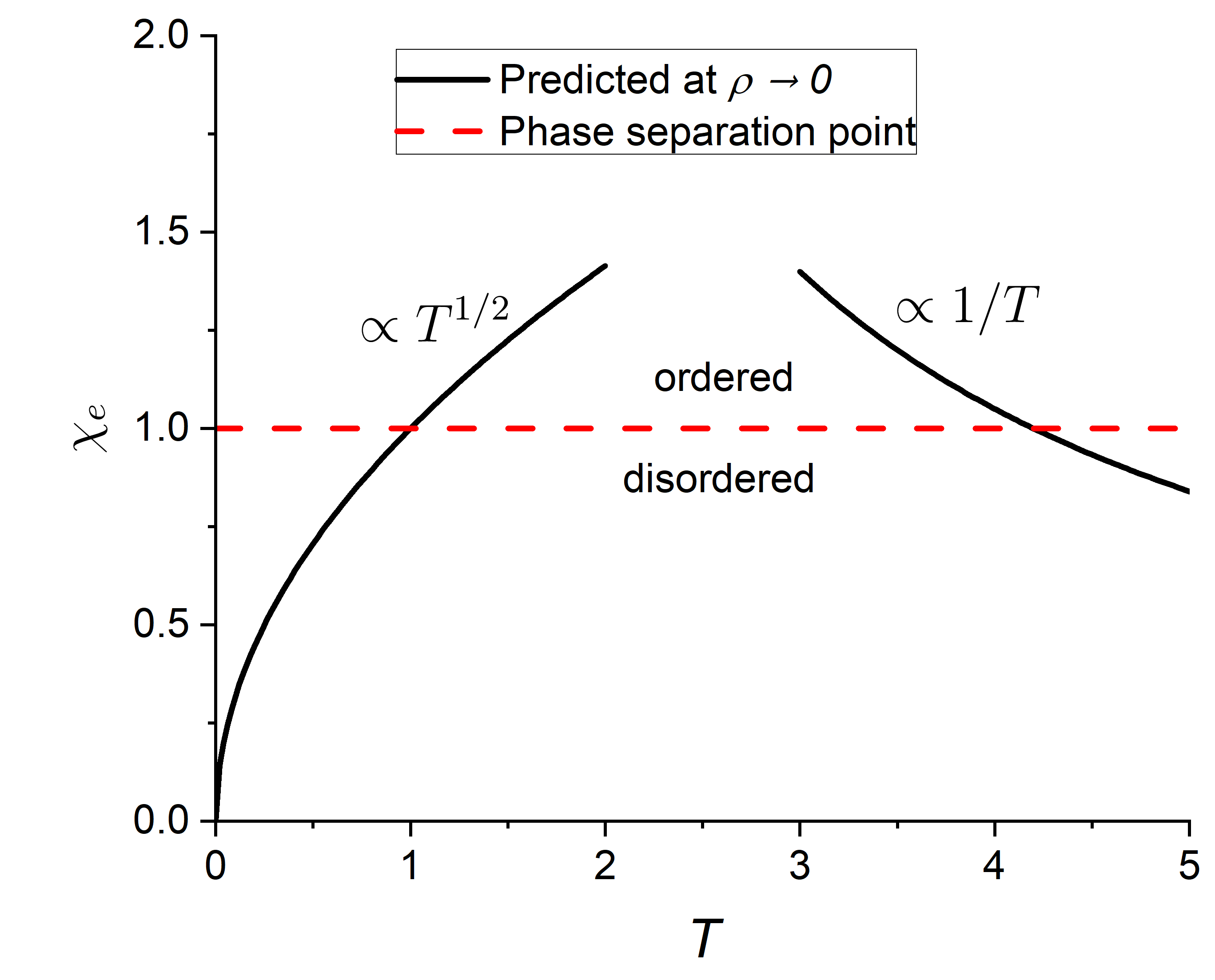}
	\caption{}
	\end{subfigure}
 \caption{Predicted $\chi_e(T)$ dependencies (black lines) for the studied CG models. $\chi_e$ at the phase separation transition was set to unity (red lines). (a) Models with high monomer number density $\rho$. The $z_\infty\alpha/k_B$ prefactor was set to unity. (b) Models with $\rho\to0$ (for concreteness, non-truncated $U_\text{nonbonded}$ is assumed to be minimized at $r=r_{cut}$, so $m=2$). Left and right black curves correspond to the $T\ll a_{xx}/k_B$ and $T\gg a_{xx}/k_B$ regimes, respectively.}
  \label{fig:1}
\end{figure}

In addition, the linear $\chi_e=z_\infty\alpha/(k_BT)$ approximation (Eq. \ref{eq1} truncated at $n=1$) almost perfectly coincides with the full nonlinear $\chi_e$ in Eq. \ref{eq1} if a CG model has a high $\rho$ (see refs. \cite{petrov2026universality,petrov2024simple}). Therefore, $\chi_e\propto 1/T$ for such models, and the ordinary UCST phase behavior is expected (Fig. \ref{fig:1}a).

Let us review the opposite case of low-$\rho$ models. As $\rho\to 0$, the model melt starts to resemble a gas of CG monomers at the length scale of one monomer. As a first approximation, we can neglect the $r$- and $T$-dependence of the correlation hole in low-density homopolymer melts \cite{taylor1995born} and express the potential of mean force between two inter-chain neighbors just as the potential energy of nonbonded interaction: $w(r)\approx U_\text{nonbonded}(r)=a_{ij}v(r_{ij})$. This approximation is equivalent to assuming that the "reference" homogeneous polymer melt with infinitely long chains closely resembles a gas of disconnected CG monomers near $r_{cut}$. As a result, the distribution of distances between two inter-chain neighbors in the "reference" state becomes in this approximation: $P(r)\propto r^2\exp[-a_{xx}v(r)/(k_BT)]$. First, let us review the high-$T$ limit ($T\gg a_{xx}/k_B$). In this approximation, $P(r)$ tends to a $T$-independent form $P(r)\propto r^2$ similar to the high-$\rho$ regime described above. This is expected since at $\rho\to0$ and $T\to\infty$, the system is expected to resemble an ideal gas of CG monomers, which has the potential of mean force $w(r)\approx 0$ similar to the high-$\rho$ mean-field regime. Therefore, the distributions of $v$ and $k_{neigh}$ become $T$-independent as $T\gg a_{xx}/k_B$ in the $\rho\to 0$ models, which leads to a $\chi_e(T)$ dependence similar to Eq. \ref{eq5}. Finally, at high $T$, the nonlinear ($n\geq 2$) terms in Eq. \ref{eq5} are small, and $\chi_e$ is $\chi_e\propto 1/T$ similarly to the $\rho\to\infty$ case (see rightmost curve in Fig. \ref{fig:1}b).

Let us focus now on the low-$T$ limit ($T\ll a_{xx}/k_B$) in the $\rho\to 0$ models. In this approximation, the other-chain neighbors of a selected monomer will be situated mostly near the cutoff radius of the interaction $r_{cut}$, since $v(r)$ is purely repulsive. Therefore, we can Taylor-expand $v(r)$ near $r_{cut}$, and $v(r)$ will take the following form to the leading order in $r-r_{cut}$ for the majority of neighbors: $v(r)\approx v^{(m)}(r_{cut})(r-r_{cut})^m/m!$, where $m$ is the smallest integer at which the $m$-th derivative of $v(r)$ at $r=r_{cut}$ is nonzero. It is worth mentioning that for the majority of CG models, $m=2$, since non-truncated $v(r)$ attains a minimum at $r=r_{cut}$; moreover, for the conservative potential used in DPD, $v(r)=(r-r_{cut})^2/2$, and, therefore, the above approximate expression for $v(r)$ is exact for the models using DPD-like nonbonded potential.

This expression for $v(r)$ allows the inversion of the (monotonic) $v(r)$ function: $r(v)\approx r_{cut}-v^{1/m}C_m^{-1/m}$, where $C_m\equiv (-1)^mv^{(m)}(r_{cut})/m!>0$. By changing the variables in $P(r)$, we arrive at the following expression for $P(v)$: $P(v)\propto \exp[-a_{xx}v/(k_BT)]v^{-1+1/m}(1+c_1v^{1/m}+c_2v^{2/m})$, where $c_1$ and $c_2$ are $v$-independent constants. Since $v(r_{cut})=0$ and since the majority of other-chain neighbors are situated at distances close to $r_{cut}$ from the selected monomer, $v$ is small for the majority of neighbors. Hence, we can retain the leading term in $v$ in the above expression for $P(v)$ and obtain the following result: for low-$\rho$ CG models at low enough $T$ ($a_{xx}\gg k_BT$), the distribution of $v$ is approximately a gamma distribution: $P(v)\propto v^{-1+1/m}\exp[-a_{xx}v/(k_BT)]$. Therefore, the $n$-th cumulant of $v$ will be proportional to $T^n$ when $\rho\to0$ and $T\ll a_{xx}/k_B$. In particular, $\langle v\rangle\approx k_BT/(a_{xx}m)\propto T$ in this limit, where brackets denote averaging in the "reference" state of a model (this result was also obtained in ref. \cite{petrov2026universality}; however, that work focused on a crude assessment of the dependence of $\kappa_n\left[z\right]_{0,N\to\infty}$ on other CG model parameters at fixed temperature, and thus did not capture the $T$-dependence of these coefficients (including $z_\infty$) qualitatively correctly at any $\rho$).

Crucially, $\langle k_{neigh}\rangle$ also depends on $T$. $\langle k_{neigh}\rangle\propto \int_0^{r_{cut}}P(r)dr$, where $P(r)\propto r^2\exp[-a_{xx}v(r)/(k_BT)]\approx r^2\exp[-a_{xx}v^{(m)}(r_{cut})(r-r_{cut})^m/(m!k_BT)]$ is the probability density of finding an other-chain neighbor at a distance $r$ in the considered limit ($\rho\to 0$, $T\ll a_{xx}/k_B$). Upon changing integration variable $r$ to $x\equiv r_{cut}-r$ and noticing that the neighbors with $x\approx 0$ mostly contribute to the integral in the low-$T$ limit, we can write $\langle k_{neigh}\rangle\propto \int_0^\infty (r_{cut}-x)^2\exp[-a_{xx}C_mx^m/k_BT)]dx$. As a result, to the leading term in $k_BT/a_{xx}$, $\langle k_{neigh}\rangle$ will be equal to $\langle k_{neigh}\rangle\propto (k_BT/a_{xx}C_m)^{1/m}\propto T^{1/m}$ in the $\rho\to 0$, $T\ll a_{xx}/k_B$ limit.

Let us now construct the $\chi_e(T)$ dependence in the $T\ll a_{xx}/k_B$ limit of $\rho\to 0$ CG models. In these models, the absolute value of the $O(\alpha^2)$ remainder of the sum in Eq. \ref{eq1} is smaller than the linear term $\chi_e^{(1)}=z_\infty\alpha/(k_BT)$ but is nonnegligible. However, one can use Eq. \ref{eq2} and the theory developed above for the distribution of $v$ and $k_{neigh}$ to show that the cumulants of $z$ can be represented as $\kappa_n[z]_{0,N\to\infty}\approx \sum_i A_i T^{x_i}$, where $A_i$ are $T$-independent constants and $x_i>0$. Rewriting Eq. \ref{eq1} as $\chi_e\approx \chi_e^{(1)}+\sum_{n=2}^\infty\sum_i B_iT^{x_i-n}$ and using the low-$T$ limit, one can demonstrate that the aforementioned fact that $\chi_e^{(1)}>\sum_{n=2}^\infty\sum_i |B_i|T^{x_i-n}$ is sufficient to prove that the sign of the $\partial \chi_e/\partial T$ equals to the sign of $\partial \chi_e^{(1)}/\partial T$. As a result, the overall shape of the $\chi_e(T)$ function is determined by the shape of $\chi_e^{(1)}(T)$ only (as confirmed in simulations in the next section). Therefore, $\chi_e(T)\propto z_\infty(T)\alpha/(k_BT)$; assuming the independence of the random variables $k_{neigh}$ and $v(r)$, one can write $z_\infty\approx \langle k_{neigh}\rangle \langle v\rangle\propto T^{1+1/m}$. Hence, $\chi_e$ is an increasing power-law function of $T$ in the $T\ll a_{xx}/k_B$ limit of low-$\rho$ CG models: $\chi_e(T)\propto T^{1/m}$. For $U_\text{nonbonded}$ that yield $m=2$, $\chi_e$ is $\chi_e\propto T^{1/2}$ (see leftmost curve in Fig. \ref{fig:1}b).

As a side note, one can obtain the aforementioned result for the special case of DPD models by noticing that in the $\rho\to 0$ limit, two-monomer correlations must be a dominant contribution to $\chi_e$. On the other hand, Eq. \ref{eq1} reduces to a scaled difference of the second virial coefficients measured in gases with interparticle interaction potentials $U=a_{AB}v(r)$ and $U=a_{xx}v(r)$ (Eq. S24 in ref. \cite{petrov2026universality}) if only two-monomer correlations are taken into account and, as mentioned above, the correlation hole does not influence strongly the $r$- and $T$-dependence of the potential of mean force in the "reference" state. In turn, ref. \cite{minkara2019new} calculated the second virial coefficients for a gas of disconnected monomers in DPD; by taking the $T\ll a_{xx}/k_B$ limit in their Eq. 20, one can show that the difference of second virial coefficients yields $\chi_e\propto T^{1/2}$, as predicted above for the particular case of $m=2$.

To sum up, the $T$-dependence of $\chi_e$ strongly depends on the monomer number density set in the studied CG models. If $\rho$ is high, $\chi_e\propto 1/T$ is a monotonically decreasing function of $T$, which leads to the UCST phase behavior of such models. On the other hand, low-$\rho$ models have $\chi_e$ increasing with $T$ as a power law ($\chi_e\propto T^{1/m}$ where $m$ is an integer) at $T\ll a_{xx}/k_B$ and decreasing with $T$ as $\chi_e\propto 1/T$ at $T\gg a_{xx}/k_B$. As a result, by selecting appropriately high mismatch energy $\alpha$, one can observe an LCST in the low-$\rho$ models having any compositional symmetry and any chain architecture, including polymer blends and block copolymer melts (Fig. \ref{fig:1}b). As temperature rises to the high-$T$ regime, $\chi_e$ decreases with $T$, which yields the appearance of UCST point above the LCST, leading to the closed-loop phase behavior (Fig. \ref{fig:1}b). Importantly, our theory predicts the existence of the closed-loop phase behavior for a CG model of any multicomponent polymer melt with sufficiently low $\rho$ and high mismatch energy $\alpha$; however, the UCST point in this reentrant transition might be very large for some models, which might prevent one from observing it in simulations.  Moreover, the above derivations were not polymer-specific; therefore, we also expect the same mechanism to cause similar temperature response of the CG models of simple fluids with the studied purely repulsive interaction potentials.

\section{Molecular Dynamics Simulations and Comparison to Theory}

\subsection{Methods}

To test the theoretical predictions, we constructed high- and low-$\rho$ CG models of diBCP melts and binary blends. We made the models fully symmetric to remove the compressibility-related effects stemming from the difference between pure species $A$ and $B$ and to make the species structurally identical as assumed in the theory. We represented monomers as point-like particles; bonds between all monomers were modeled using harmonic potential $U_\text{bonded}^{ij}=K(r_{ij}-r_0)^2/2$, where $r_{ij}$ is the distance between pair of bonded monomers $i$ and $j$, $K$ and $r_0$ are constants. For all studied models, we chose to model stiff bonds ($K=100$); as a result, the $b(T)$ dependence was weak, and $\bar{N}=N\rho^2b^6$ was also a weak function of $T$ and was equal to $\bar{N}\approx 800$ for all studied models, which is within the $\bar{N}\gtrsim 300$ range in which the universality hypothesis was shown to work \cite{petrov2026universality,glaser2014universality}. All monomers were interacting through pairwise repulsive nonbonded potential used in DPD: $U_\text{nonbonded}^{ij}=a_{ij}(1-r_{ij})^2/2$ for $r_{ij}<1$ and $U_\text{nonbonded}^{ij}=0$ otherwise. Here, $a_{ij}$ are the coefficients depending on the type of interacting monomers \cite{groot1997dissipative}. The like-type interaction coefficients did not depend on the particle type: $a_{AA}=a_{BB}\equiv a_{xx}$; this, combined with the type-independent bond potential, made all monomers structurally identical. In the diBCP melt models, we set the fraction of $A$ monomers in a diBCP chain to $f=0.5$; in the binary blend models, the fraction of $A$-type chains was equal to $\phi=0.5$. In the diBCP models, the simulation box size was chosen to accommodate more than three full lamellar periods in order to make finite box size effects negligible \cite{petrov2024simple}; for the same reason, in binary blend models, the box side length was several times bigger than chain radius of gyration \cite{rumyantsev2026scaling}. Simulations were performed using LAMMPS software in NVT ensemble using Nose-Hoover thermostat. For each model, we varied temperature $T$ in the range $T\in[0.5,5]$; as shown in refs. \cite{zhu2011phase,sun2014structural}, $T=0.5$ is above the crystallization and vitrification temperatures for DPD models of liquids with any $\rho$ (if $a_{xx}\leq 75/\rho$ as in this work).

During the investigation of the phase behavior of models, we varied $T$ at a fixed value of $\alpha>0$ calculated as follows. In the binary blend models, $\chi_eN$ at the critical point was assumed to be close to the mean-field Flory-Huggins theory prediction $\chi_e^{crit}N=2$ (as shown in ref. \cite{willis2019calibration}, non-mean field finite-$\bar{N}$ effects affect $\chi_e^{crit}N$ weakly in binary blends). We set $\alpha$ in the binary blend models to the value $\alpha_{crit}(T=1)$ that renders Eq. \ref{eq1} equal to $\chi_e^{crit}=2/N$ at $T=1$. Similarly, diBCP models undergo order-disorder transition (ODT) at $\chi_e^{ODT}N=10.495+41.0\bar{N}^{-1/3}$, which was predicted theoretically by the Brazovskii-Fredrickson-Helfand (BFH) theory \cite{fredrickson1987fluctuation,brazovskii1975phase} and recently established in the simulations of CG diBCP melt models if the definition of $\chi_e$ in Eq. \ref{eq1} is adopted \cite{petrov2026universality}. Therefore, since $\bar{N}\approx 800$ in all studied models, $\alpha$ in diBCP melt models was set to $\alpha_{ODT}(T=1)$, which is the value realizing the equality $\chi_e(T=1)=\chi_e^{ODT}=(10.495+41\bar{N}^{-1/3})/N\approx 14.9/N$.

In order to characterize the phase separation transition quantitatively, we measured the normalized peak scattering intensity $S(q_{peak})/(\rho N)$. $S(q)$ was calculated as the Fourier transform of composition fluctuations correlation function \cite{petrov2024simple}. The peak of $S(q)$ characterizes how strongly species are segregated in the system: higher $S(q_{peak})$ corresponds to a more phase separated material (i.e., $S(q_{peak})$ is directly related to $\chi_e$ at a fixed $\bar{N}$ and chain architecture). For the diBCP melt model, the peak intensity was established by fitting a smooth function near the $S(q)$ peak as described in ref. \cite{petrov2024simple}. For the binary blend model, peak of $S(q)$ was measured at $q=2\pi/L_{box}$, where $L_{box}$ is the cubic simulation box side length (the largest length scale in a simulation). Measurements of the peak scattering intensity were performed from two initial states: fully homogeneous and well-ordered; this was done at all $T$ except for $T$ at which spontaneous ordering from a homogeneous state to a well-defined ordered structure was observed. Therefore, the errors of the plotted peak scattering intensity stemmed from the difference between the simulations initialized by a homogeneous and an ordered state.

\subsection{Simulation Results}

\begin{figure}[!t]
  \centering
  \includegraphics[width=\linewidth]{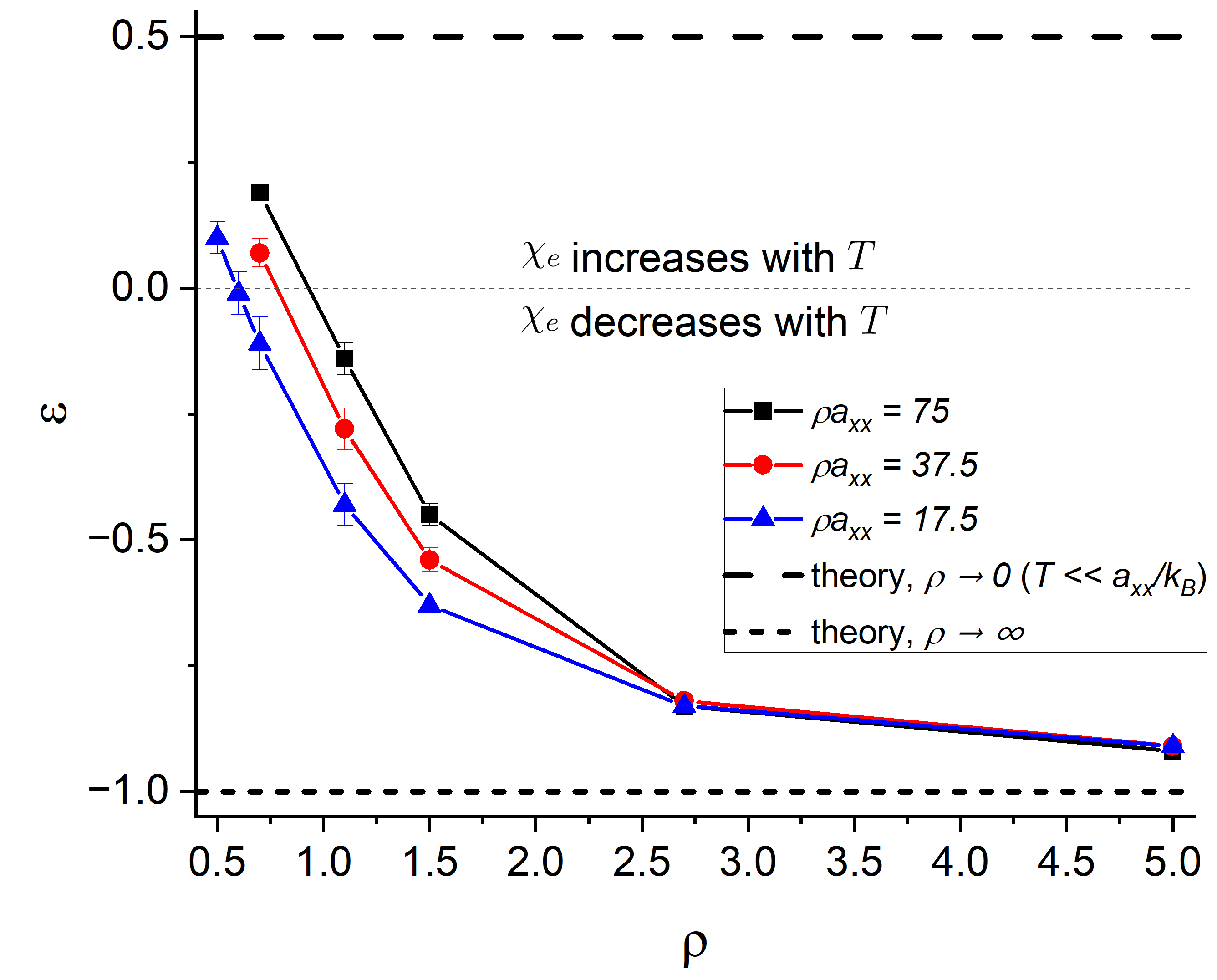}
  \caption{$\rho$-dependence of the exponent $\epsilon$ characterizing the $T$-dependence of the linear approximation for $\chi_e$ ($\chi_e^{(1)}=z_\infty(T)\alpha/(k_BT)\propto T^\epsilon$) at different values of $\rho a_{xx}$. $r_0=1$ was used in all models in this plot. Error bars are errors of the $z_\infty(T)$ power law fitting only (error of $z_\infty$ measurement was small).}
  \label{fig:zinf_exponents}
\end{figure}

Our first objective was to understand at what monomer densities $\rho$ the predicted "small-$\rho$" and "large-$\rho$" effects can be observed in a simulation. We constructed a range of models having $\rho$ values $\rho\in[0.5,5]$ typically used in DPD \cite{glaser2014universality,groot1997dissipative}. For each $\rho$, we also set three different values of $a_{xx}$: $a_{xx}=\{17.5/\rho,37.5/\rho,75/\rho\}$ (or, equivalently, we set three different dimensionless system compressibilities $\kappa^{-1}$ at each $\rho$ at a fixed $T$, since $\kappa^{-1}\propto a_{xx}\rho/(k_BT)$; the choice $a_{xx}=75/\rho$ at $T=1$ yielded the compressibility of water \cite{groot1997dissipative}). After that, we performed the simulations of the homogeneous state ($\alpha=0$) of each of these models and measured $z_\infty$ values according to Eq. \ref{eq3}. As shown in Fig. S2, $z_\infty(T)$ formed a curve close to a straight line in log-log coordinates for all studied models \cite{supplemental}. Therefore, as a first crude assessment, we assumed that a simple power-law $z_\infty(T)$ dependence holds in the simulated models. This allowed us to define the exponent $\epsilon$ in the $T$-dependence of the linear approximation for $\chi_e$: $\chi_e^{(1)}=z_\infty(T)\alpha/(k_BT)\propto T^\epsilon$. As shown theoretically in Section II and in simulations below, the dominant $\chi_e^{(1)}$ term fully determines the $T$-dependence of $\chi_e$ in all studied models. As a result, $\epsilon$ characterizes how $\chi_e$ depends on $T$ in different models: $\epsilon>0$ indicates a monotonically increasing $\chi_e(T)$ function (and vice versa). Fig. \ref{fig:zinf_exponents} shows the $\epsilon(\rho)$ dependence.

First, we observed that as $\rho$ increased past $\rho\approx2.7$, $\epsilon$ values started to saturate and approach the $\epsilon\approx -1$ limit expected in the mean-field regime. Moreover, at $\rho\geq2.7$, $\epsilon$ did not depend on the $a_{xx}\rho$ parameter characterizing compressibility at a fixed $T$. Thus, at these monomer densities, local liquid structure was weakly dependent on $T$ and other microscopic model parameters except for $\rho$, which additionally signified the onset of the mean-field regime. Therefore, we can consider $\rho=5$ as the candidate "high-$\rho$" regime density in which near-mean-field behavior could be potentially pronounced. On the other hand, as $\rho$ decreased below $\rho\approx 2.7$, $z_\infty$ acquired a strong $T$-dependence that resulted in a sharp increase of $\epsilon$ as predicted theoretically in the $\rho\to0$, $T\ll a_{xx}/k_B$ limit. It is worth mentioning that the temperature values $T\in[0.5,5]$ used to determine $\epsilon$ satisfied the $T\ll a_{xx}/k_B$ condition for any model in Fig. \ref{fig:zinf_exponents} having $\rho\lesssim1$ ($k_B=1$ in our units). Importantly, below the crossover density $\rho\approx 1.0$, $\epsilon$ exceeded zero, indicating the appearance of a \textit{growing} $\chi_e^{(1)}(T)$ function and, therefore, potential LCST/closed-loop phase behavior of these models. Therefore, we selected $\rho=0.7$ as the density at which "low-$\rho$" effects predicted in Section II can be probed in a simulation. Finally, more compressible systems have smaller $a_{xx}$ and, therefore, "softer" repulsive potentials at a given $\rho$ and $T$; hence, these systems are expected to be closer to the mean-field behavior. Fig. \ref{fig:zinf_exponents} confirms this intuition: $\epsilon(\rho)$ deviates from the mean-field $\epsilon=-1$ prediction at a lower $\rho$ for models with smaller $a_{xx}\rho$ values.

To study the high-$\rho$ models in greater detail, we constructed one diBCP melt model and one binary blend model with $\rho=5$. All chains had degree of polymerization $N=20$, $a_{xx}=15$, and $r_0=1.0$. Fig. \ref{fig:zinf_exponents} pointed out that these models might be described well by the theory in the mean-field $\rho\to\infty$ regime. We confirmed it from several different angles. First, the $\rho\to\infty$ theory assumed the linear approximation for $\chi_e$: $\chi_e\approx z_\infty\alpha/(k_BT)$. Fig. S1a confirmed that nonlinear in $\alpha$ terms in Eq. \ref{eq1} were small for all studied $T$. Second, the theory posited that $z_\infty$ depends weakly on $T$ due to the weak $T$-dependence of the distributions of $v$ and $k_{neigh}$. This was supported by the weak $\langle v\rangle(T)$ and $\langle k_{neigh}\rangle(T)$ dependencies in Fig. S3. Since only these two dependencies determined the $z_\infty(T)$ scaling (Fig. S2a), $z_\infty$ was a weak function of $T$ as well. This, in turn, resulted in the $\chi_e\propto 1/T$ dependency (or, equivalently, $\epsilon\approx -1$, Fig. \ref{fig:zinf_exponents}). Therefore, these data demonstrated that the chosen $\rho=5$ model adhered well to the $\rho\to\infty$ theory developed in Section II.

\begin{figure*}[!t]
    \centering

    \begin{subfigure}{0.4\textwidth}
        \centering
        \includegraphics[width=\linewidth]{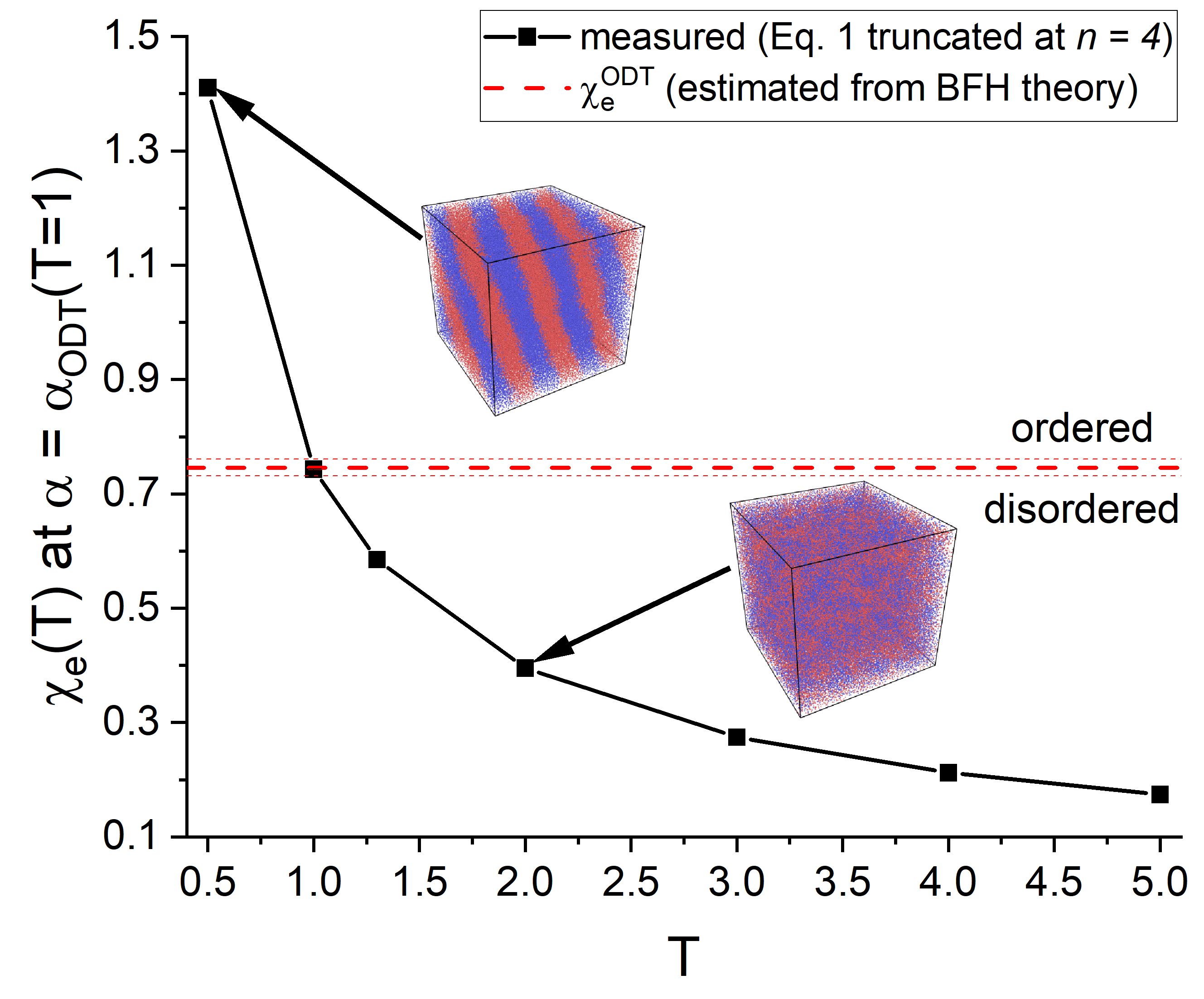}
        \caption{}
        \label{xt_proj_fig_2_BCP_chiandsnaps.png}
    \end{subfigure}
    \hfill
    \begin{subfigure}{0.4\textwidth}
        \centering
        \includegraphics[width=\linewidth]{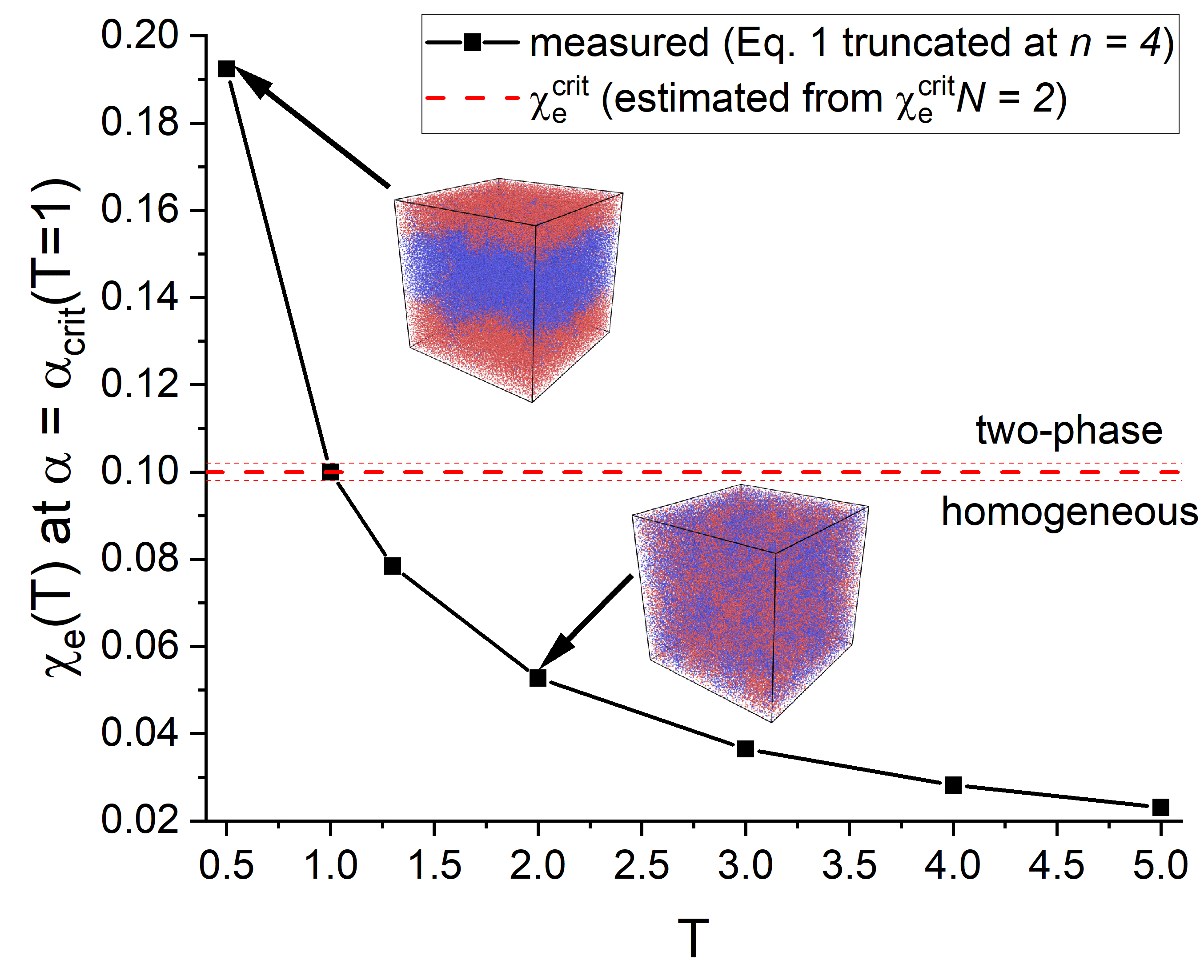}
        \caption{}
        \label{xt_proj_fig_2_blend_chiandsnaps.png}
    \end{subfigure}
    \begin{subfigure}{0.4\textwidth}
        \centering
        \includegraphics[width=\linewidth]{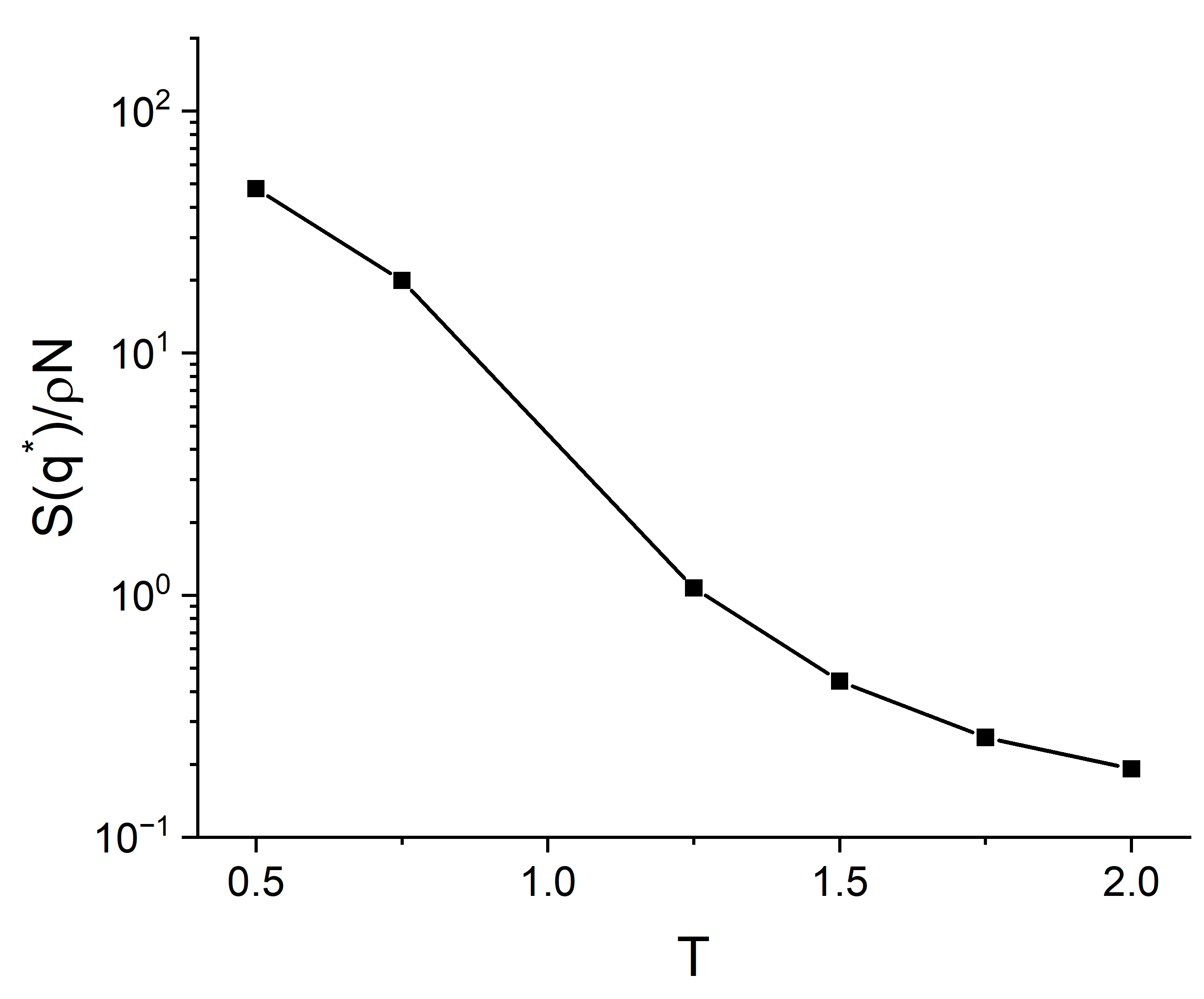}
        \caption{}
        \label{xt_proj_fig_2_BCP_sq.png}
    \end{subfigure}
    \hfill
    \begin{subfigure}{0.4\textwidth}
        \centering
        \includegraphics[width=\linewidth]{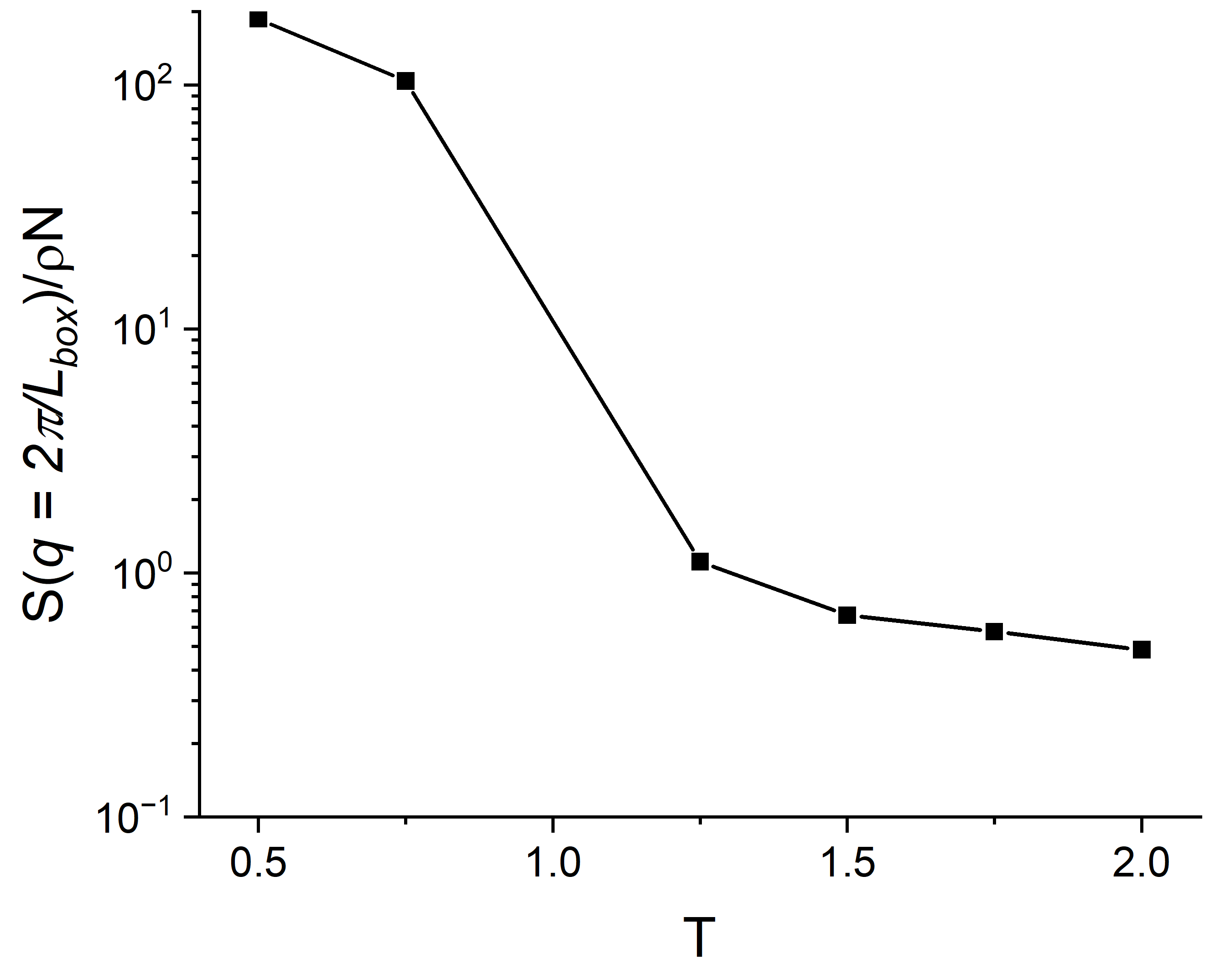}
        \caption{}
        \label{xt_proj_fig_2_blend_sq.png}
    \end{subfigure}

    \caption{UCST phase behavior in high-$\rho$ ($\rho=5$) fully symmetric models of diBCP melts (a,c) and binary blends (b,d). Black points in panels (a) and (b) show the values of $\chi_e$ estimated at each $T$ using Eq. \ref{eq1} truncated at fourth term. This truncation approximated the full Eq. \ref{eq1} well as shown in Fig. S1a. Error bars of $\chi_e$ (which were always smaller than the magnitude of the $O(\alpha^4)$ term in Eq. \ref{eq1}) are within symbol size. Mismatch energy $\alpha$ was fixed at all $T$ to $\alpha_{ODT}(T=1)$ (a,c) and to $\alpha_{crit}(T=1)$ (b,d). $\chi_e^{ODT}$ and $\chi_e^{crit}$ are shown by the horizontal thick dashed red lines. Thin dashed red lines show $\pm 2\%$ deviation from $\chi_e^{ODT}$ in (a) and from $\chi_e^{crit}$ in (b). Snapshots of ordered and disordered states are included in (a) and (b). Panels (c) and (d) show normalized peak scattering intensities at each $T$ for the diBCP melt model (c) and the binary blend model (d).}
    \label{fig:2}
\end{figure*}

Based on this prediction for the $\chi_e(T)$ function, the theory suggested the UCST phase behavior of the high-$\rho$ models. To confirm this prediction and observe phase separation, we set $\alpha$ to a fixed value ($\alpha_{ODT}(T=1)$ and $\alpha_{crit}(T=1)$ for the diBCP melt and binary blends models, respectively) calculated as described in Methods. Fig. \ref{fig:2}a,b show the $\chi_e(T)$ dependencies measured using Eq. \ref{eq1}. For both binary blend and diBCP melt models, these dependencies predicted the UCST phase behavior, which was confirmed qualitatively by the visual inspection of system snapshots (Fig. \ref{fig:2}a,b) and by the peak scattering intensity plots (Fig. \ref{fig:2}c,d), which exhibited a monotonic decrease with $T$ with a sharp drop near the transition temperature $T=1$.

After this, we studied the low-$\rho$ models having $\rho=0.7$. Fig. \ref{fig:zinf_exponents} showed that these models might be described qualitatively by the $\rho\to 0$ theory developed in Section II. To confirm this, we constructed a model of fully symmetric binary blends and a model of fully symmetric diBCP melts by setting $r_0=1.0$, $a_{xx}=107$, and $N=232$ (this $N$ was chosen to yield $\bar{N}\approx 800$ like in the $\rho=5$ models studied above). These models corresponded to the leftmost black square in Fig. \ref{fig:zinf_exponents}, having a large positive $\epsilon\approx 0.2$ that indicated potential LCST phase behavior. The $\rho\to 0$ theory relied on the assumption that the $\chi_e(T)$ dependence is mostly determined by the linear term $\chi_e^{(1)}(T)=z_\infty(T)\alpha/(k_BT)$; this was confirmed in Fig. S1b. Moreover, the small-$\rho$ theory assumed $z_\infty\approx \langle v\rangle \langle k_{neigh}\rangle$ factorization, which was supported by the data in Fig. S2b. In addition, the $\rho\to 0$ theory derived $\langle v\rangle \propto T$; the observed $\propto T^{0.88}$ scaling in Fig. S4a agreed with this prediction reasonably well. Fig. S4b demonstrated the $\langle k_{neigh}\rangle\propto T^{0.3}$ dependence, which somewhat disagreed with the theoretical $\propto T^{0.5}$ prediction presumably due to the finite density of the model. Despite the quantitative discrepancies, however, this model agreed qualitatively well with the predictions of the $\rho\to 0$ theory developed in Section II.

\begin{figure*}[!t]
    \centering

    \begin{subfigure}{0.4\textwidth}
        \centering
        \includegraphics[width=\linewidth]{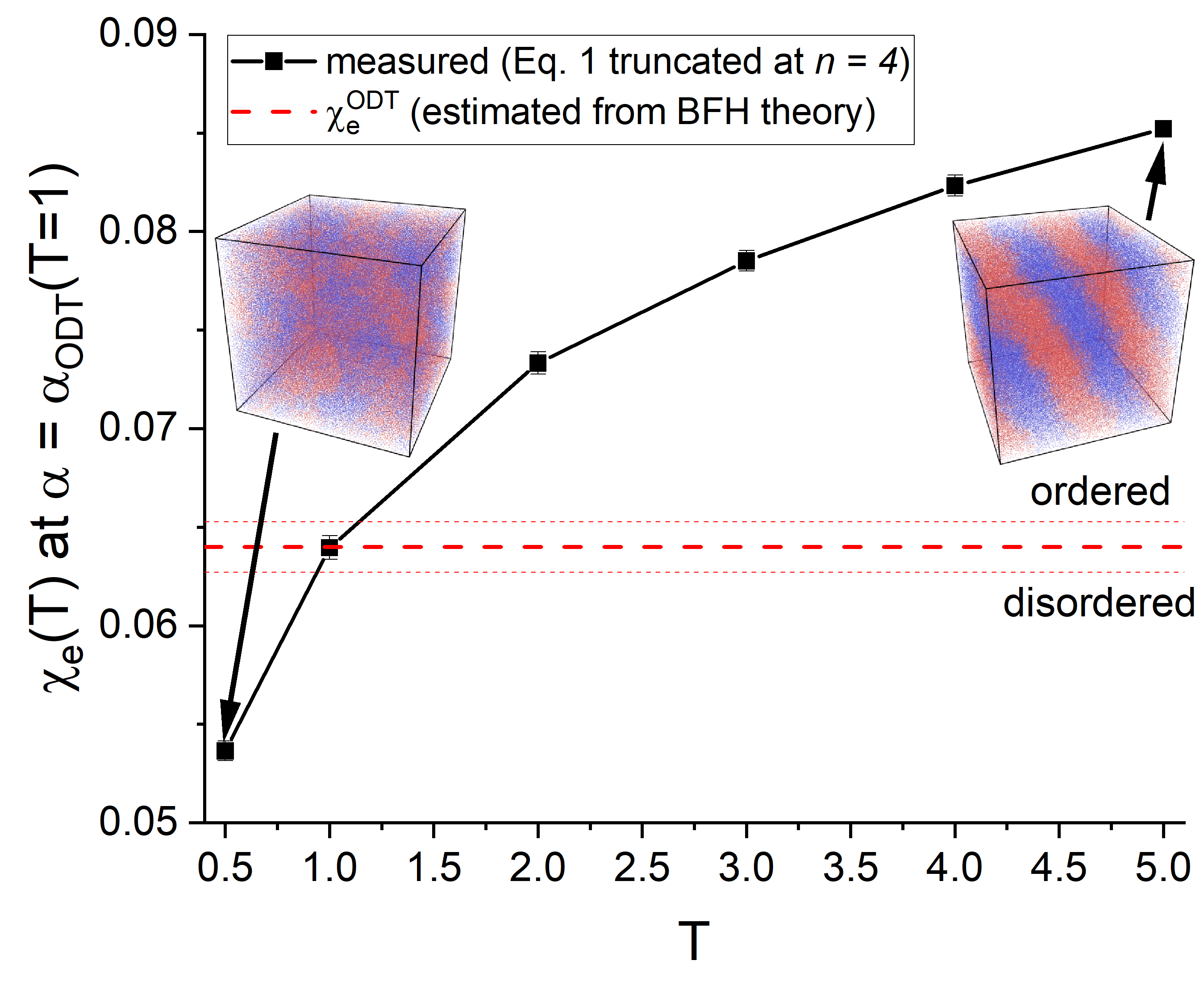}
        \caption{}
        \label{xt_proj_fig_3_BCP_chiandsnaps.png}
    \end{subfigure}
    \hfill
    \begin{subfigure}{0.4\textwidth}
        \centering
        \includegraphics[width=\linewidth]{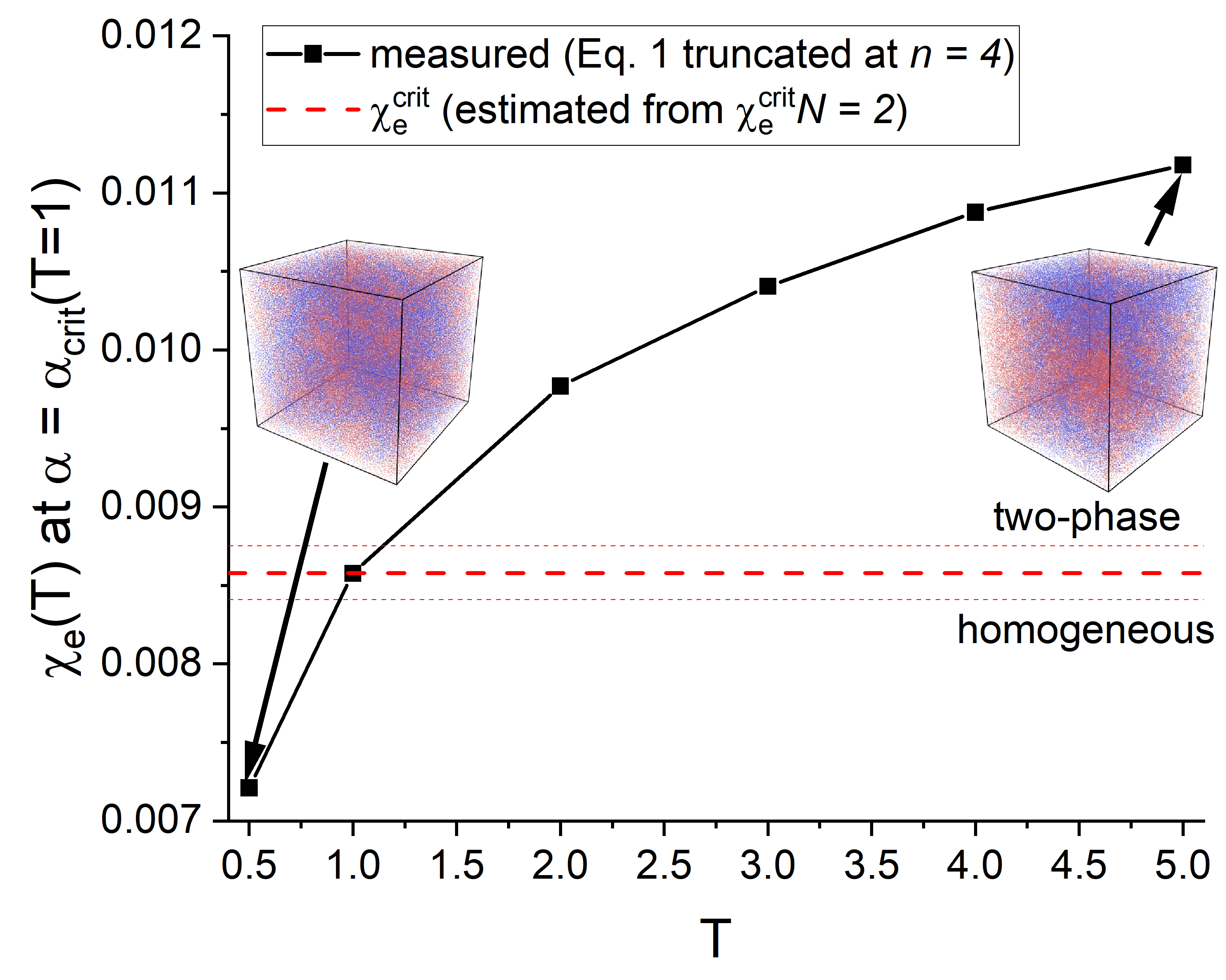}
        \caption{}
        \label{xt_proj_fig_3_blend_chiandsnaps.png}
    \end{subfigure}
    \begin{subfigure}{0.4\textwidth}
        \centering
        \includegraphics[width=\linewidth]{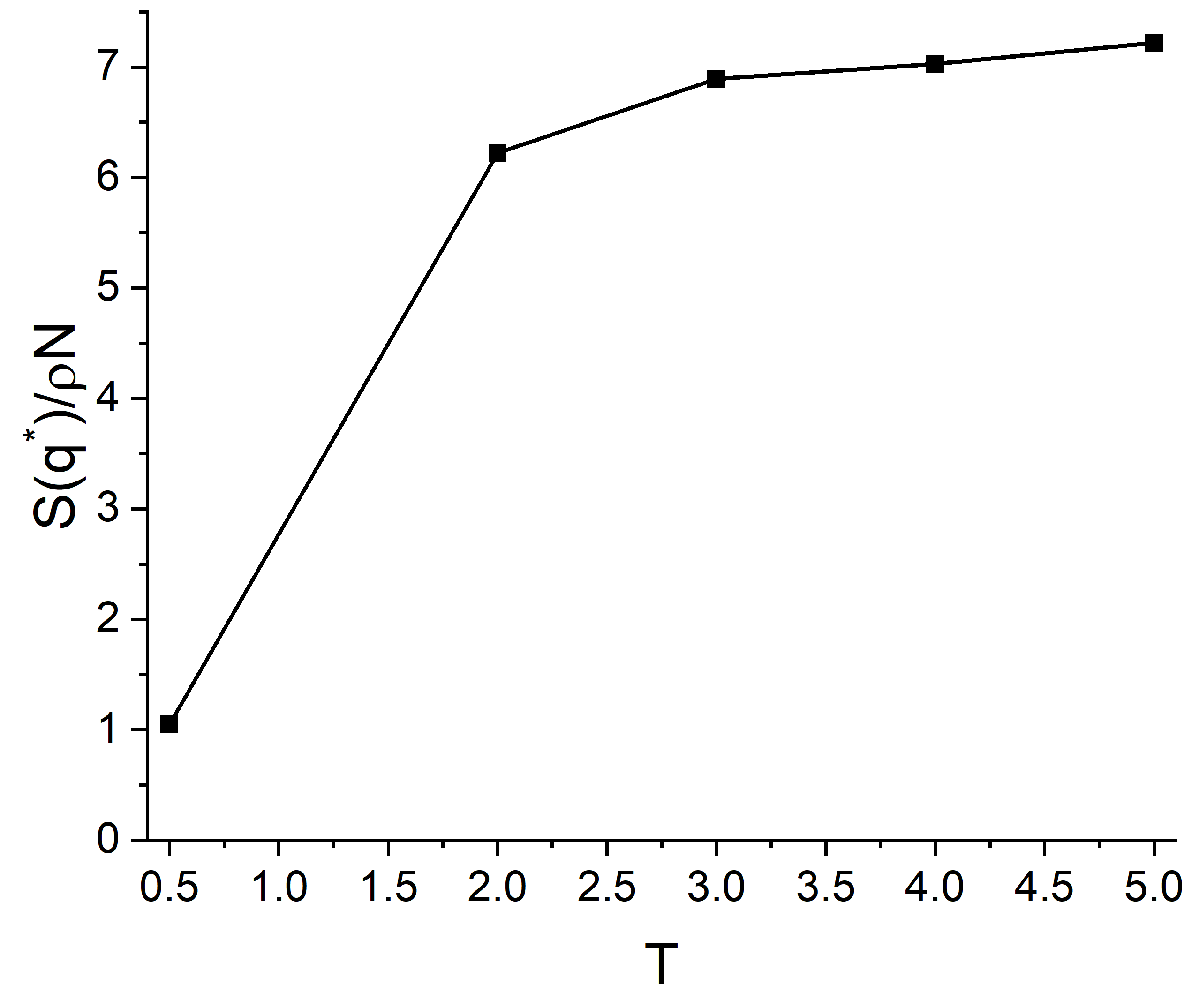}
        \caption{}
        \label{xt_proj_fig_3_BCP_sq.png}
    \end{subfigure}
    \hfill
    \begin{subfigure}{0.4\textwidth}
        \centering
        \includegraphics[width=\linewidth]{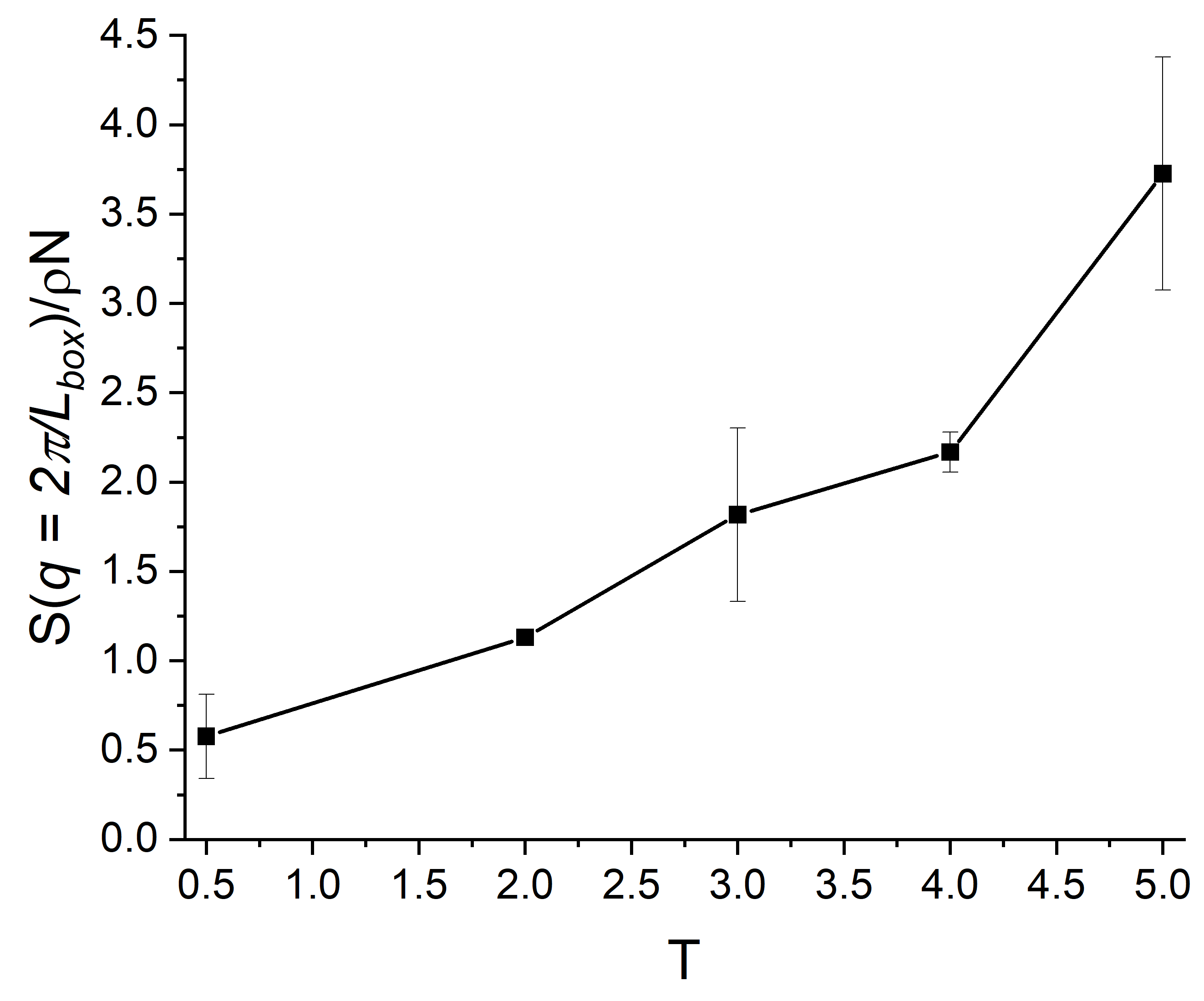}
        \caption{}
        \label{xt_proj_fig_3_blend_sq.png}
    \end{subfigure}

    \caption{LCST phase behavior in low-$\rho$ ($\rho=0.7$, $r_0=1.0$) fully symmetric models of diBCP melts (a,c) and binary blends (b,d). Black points in panels (a) and (b) show the values of $\chi_e$ estimated at each $T$ using Eq. \ref{eq1} truncated at fourth term. This truncation approximated the full Eq. \ref{eq1} well as shown in Fig. S1b. Error bars of $\chi_e$ show the maximal possible error of $\chi_e$ equal to the magnitude of the $O(\alpha^4)$ term in Eq. \ref{eq1}. Mismatch energy $\alpha$ was fixed at all $T$ to $\alpha_{ODT}(T=1)$ (a,c) and to $\alpha_{crit}(T=1)$ (b,d). $\chi_e^{ODT}$ and $\chi_e^{crit}$ are shown by the horizontal thick dashed red lines. Thin dashed red lines show $\pm 2\%$ deviation from $\chi_e^{ODT}$ in (a) and from $\chi_e^{crit}$ in (b). Snapshots of ordered and disordered states are included in (a) and (b). Panels (c) and (d) show normalized peak scattering intensities at each $T$ for the diBCP melt model (c) and the binary blend model (d).}
    \label{fig:3}
\end{figure*}

Fig. \ref{fig:3} shows the phase behavior of these models. As in the case of the $\rho=5$ models studied above, we set $\alpha$ to $\alpha_{ODT}(T=1)$ and $\alpha_{crit}(T=1)$ for the diBCP melt and binary blend models, respectively. As demonstrated in Fig. \ref{fig:3}a,b, Eq. \ref{eq1} predicted a monotonically increasing $\chi_e(T)$ function that started from $\chi_e$ well below the phase transition at $T=0.5$ and ending well above the transition point at $T\geq 2$, indicating LCST phase behavior with the lower critical ordering temperature $T=1$. Simulation data (system snapshots and the peak scattering intensity plots) confirmed this prediction, showing ordering in both diBCP melt model and binary blend model upon increasing $T$. Interestingly, we did not observe the predicted closed-loop phase behavior (i.e., the emergence of UCST point above $T=1$) presumably since the simulated $T\in[0.5,5]$ values lay well within the $T\ll a_{xx}/k_B$ range for these models; in this range, only LCST was predicted by the $\rho\to 0$ theory in agreement with the data in Fig. \ref{fig:3}. Therefore, we constructed the low-$\rho$ models of the fully symmetric binary blends and diBCP melts that are well described by the analytical theory in the $\rho\to 0$ limit and that yield, as predicted, LCST phase behavior.

Finally, we built a different pair of fully symmetric binary blend and diBCP melt models with $\rho=0.7$. In these models, we set a larger bond rest length $r_0=1.5$; $a_{xx}$ was set to $a_{xx}=107$ as in the models in Fig. \ref{fig:3}, and $N$ was set to $N=64$ to yield the same $\bar{N}\approx 800$. 
These models also qualitatively adhered to the predictions of the $\rho\to 0$ theory: $\langle v\rangle(T)$ scaling was close to linear ($\langle v\rangle\propto T^{0.84}$, Fig. S5a), $\langle k_{neigh}\rangle(T)$ increased with $T$ according to a power law (Fig. S5b), the approximation $z_\infty\approx \langle v\rangle \langle k_{neigh}\rangle$ held reasonably well (Fig. S2c), and the shape of $\chi_e^{(1)}(T)$ dependence was very close to the $\chi_e(T)$ function (Fig. S1c). However, the $\langle k_{neigh}\rangle(T)$ dependence was significantly weaker in these models compared to the models studied above in Fig. \ref{fig:3}: $\langle k_{neigh}\rangle\propto T^{0.16}$ (Fig. S5b), which led to an almost perfectly linear $z_\infty(T)$ dependence (Fig. S2c): $z_\infty\propto T^{1.00}$. Crucially, if examined more closely, the $z_\infty(T)$ dependency in Fig. S2c can be divided into two regions: with the scaling exponent slightly above $1.00$ (for $T\leq 1$) and slightly below $1.00$ (for $T\geq 2$). This, in turn, resulted in the nonmonotonic $\chi_e^{(1)}(T)=z_\infty(T)\alpha/(k_BT)$ dependency (Fig. S1c), which increased with $T$ for $T\leq 1$ and decreased with $T$ for $T\geq 2$. This led to a nonmonotonic $\chi_e(T)$ dependence as well (Fig. S1c).

\begin{figure*}[!t]
    \centering

    \begin{subfigure}{0.4\textwidth}
        \centering
        \includegraphics[width=\linewidth]{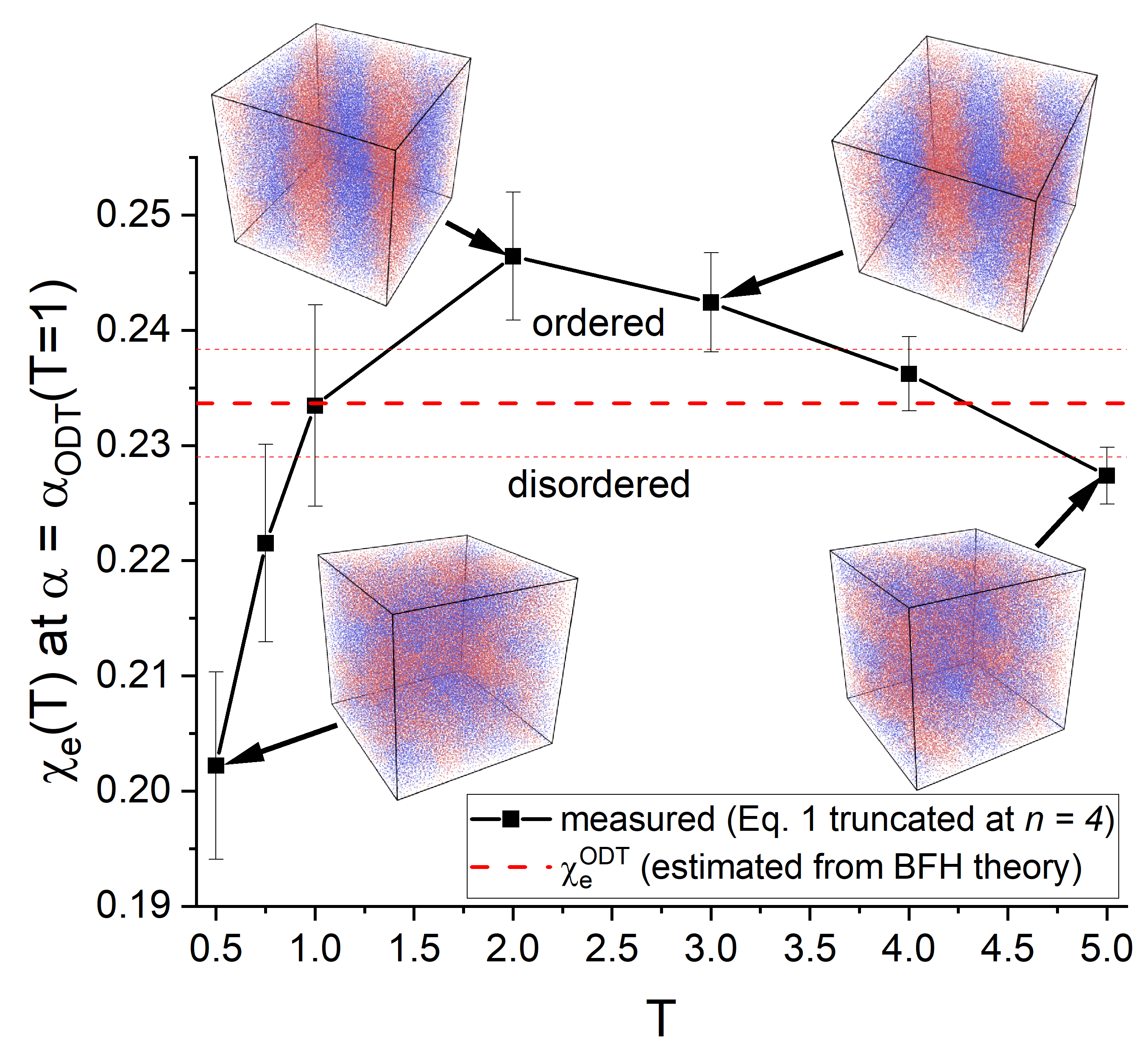}
        \caption{}
        \label{xt_proj_fig_4_BCP_chiandsnaps.png}
    \end{subfigure}
    \hfill
    \begin{subfigure}{0.4\textwidth}
        \centering
        \includegraphics[width=\linewidth]{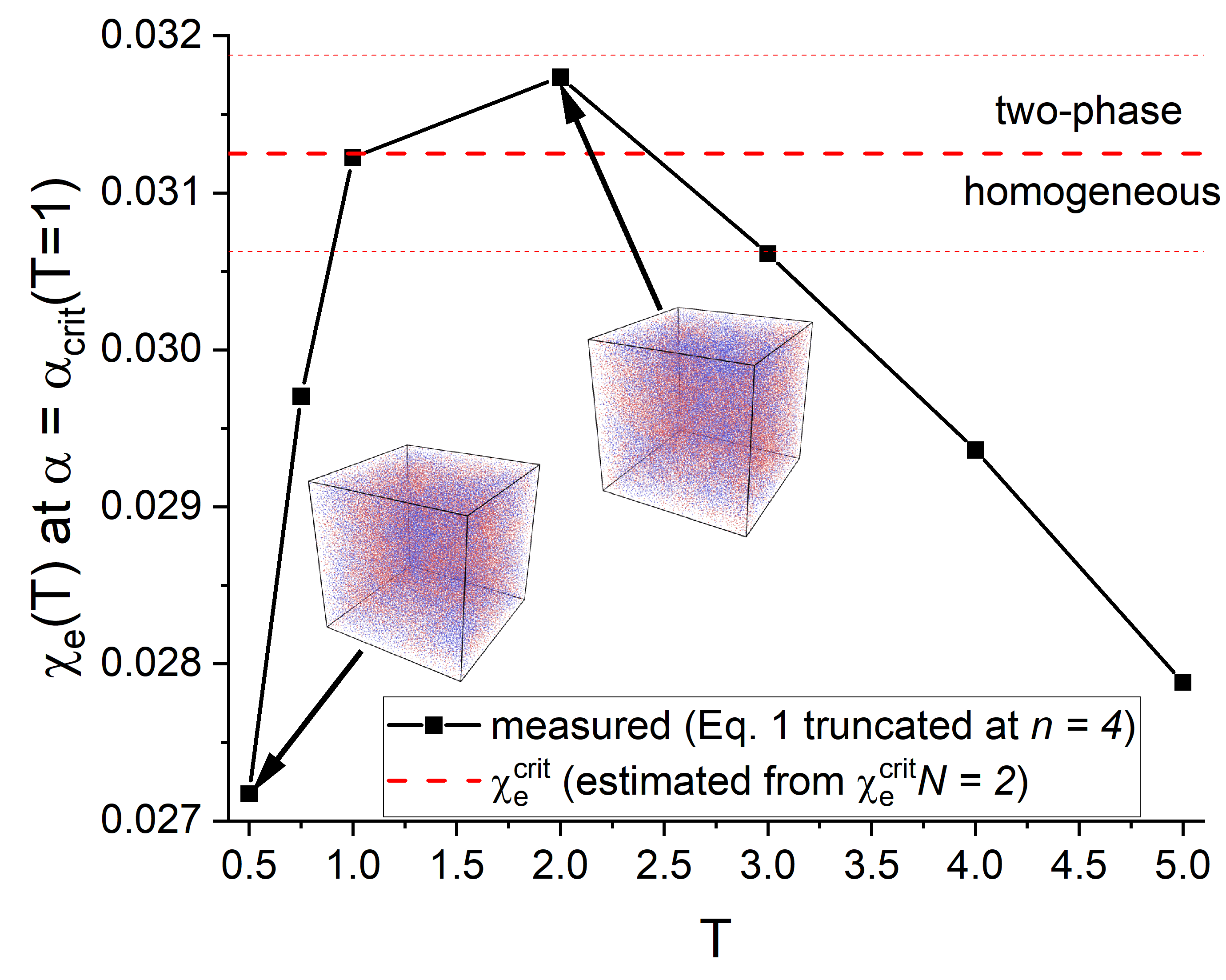}
        \caption{}
        \label{xt_proj_fig_4_blend_chiandsnaps.png}
    \end{subfigure}
    \begin{subfigure}{0.4\textwidth}
        \centering
        \includegraphics[width=\linewidth]{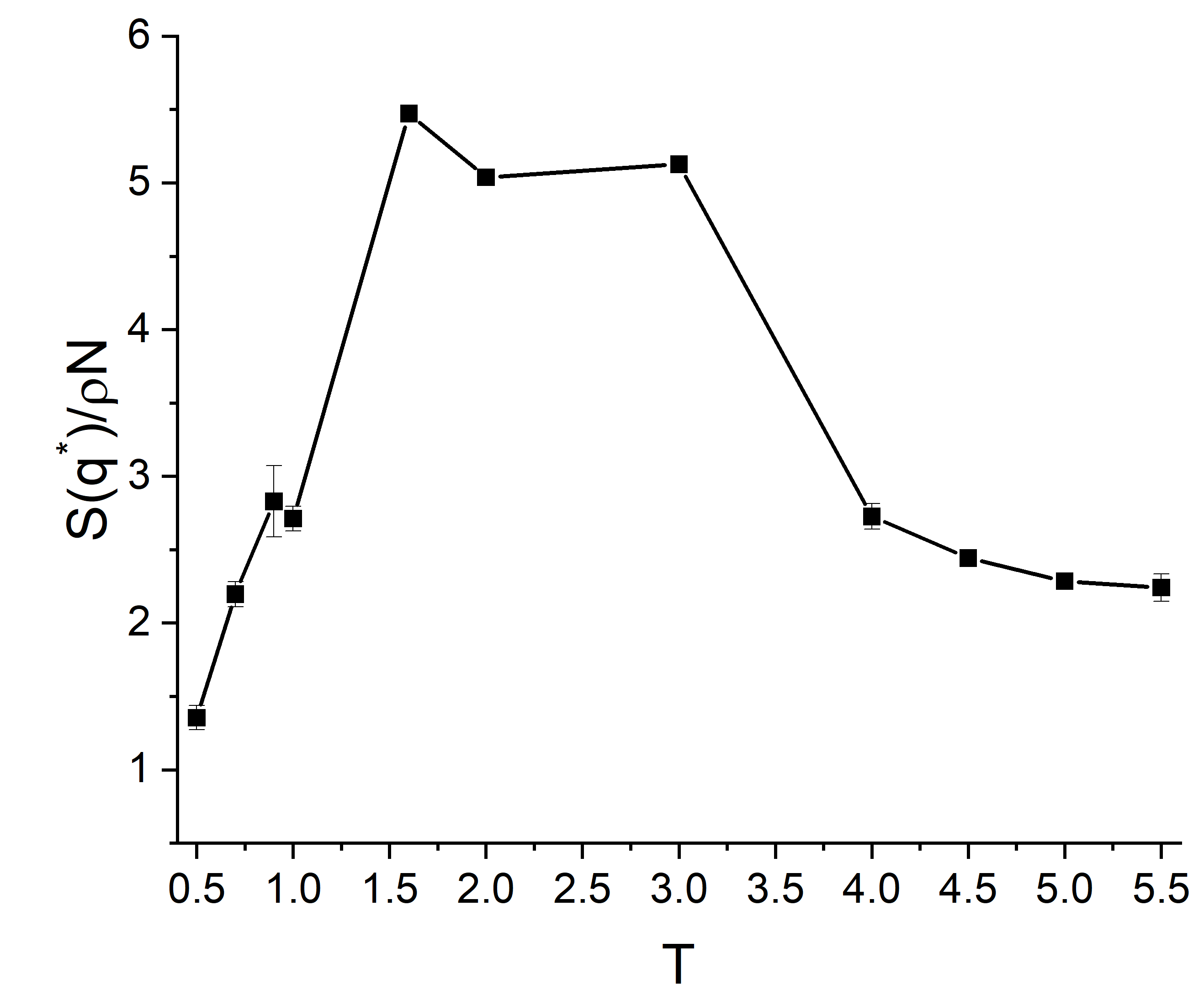}
        \caption{}
        \label{xt_proj_fig_4_BCP_sq.png}
    \end{subfigure}
    \hfill
    \begin{subfigure}{0.4\textwidth}
        \centering
        \includegraphics[width=\linewidth]{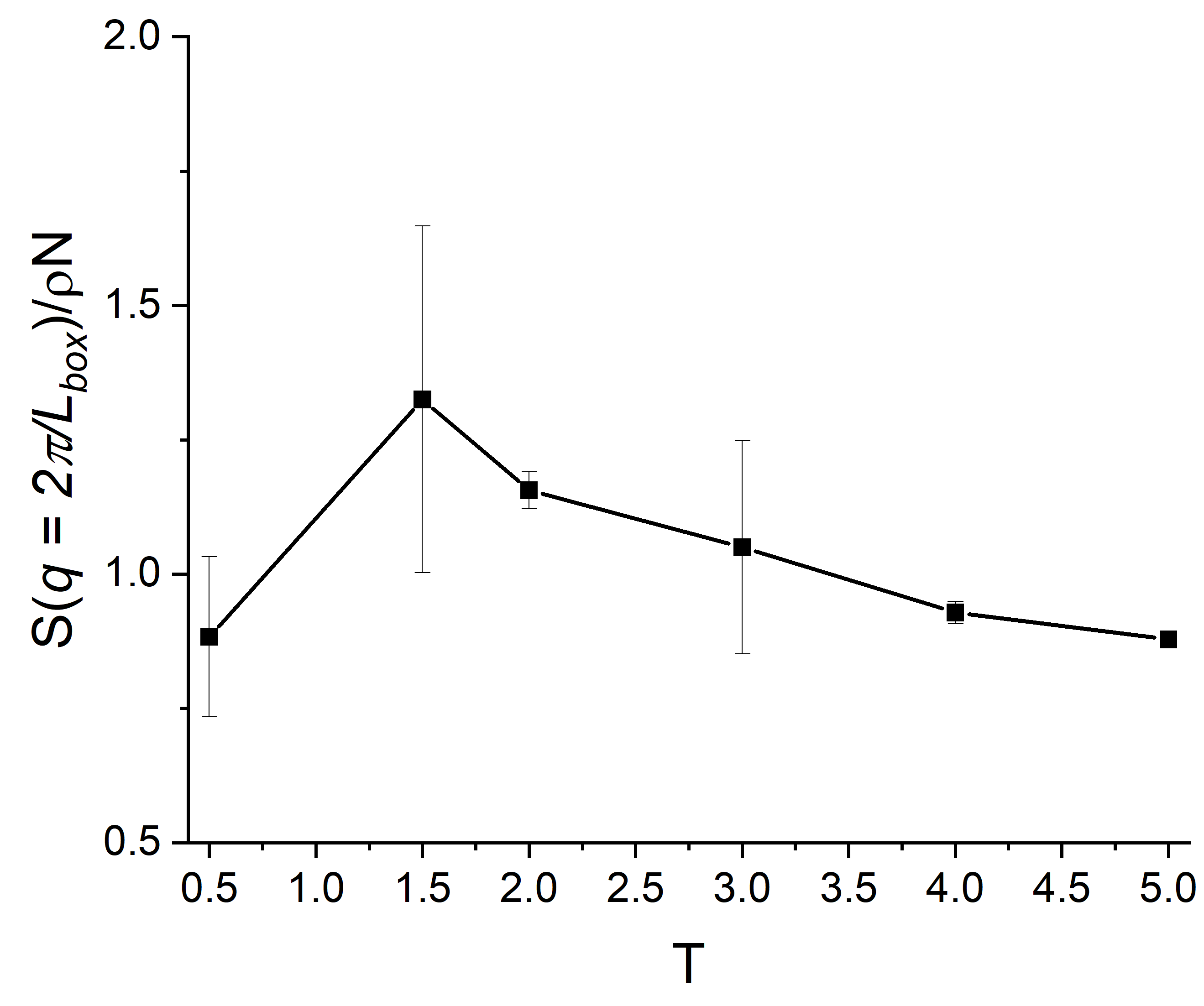}
        \caption{}
        \label{xt_proj_fig_4_blend_sq.png}
    \end{subfigure}

    \caption{Closed-loop phase behavior in a low-$\rho$ ($\rho=0.7$, $r_0=1.5$) fully symmetric model of diBCP melts (a,c). Panels (b,d) show the predicted and observed absence of notable phase separation transition in a low-$\rho$ ($\rho=0.7$, $r_0=1.5$) fully symmetric model of binary blends. Black points in panels (a) and (b) show the values of $\chi_e$ estimated at each $T$ using Eq. \ref{eq1} truncated at fourth term. This truncation approximated the full Eq. \ref{eq1} well as shown in Fig. S1c. Error bars of $\chi_e$ show the maximal possible error of $\chi_e$ equal to the magnitude of the $O(\alpha^4)$ term in Eq. \ref{eq1}. Mismatch energy $\alpha$ was fixed at all $T$ to $\alpha_{ODT}(T=1)$ (a,c) and to $\alpha_{crit}(T=1)$ (b,d). $\chi_e^{ODT}$ and $\chi_e^{crit}$ are shown by the horizontal thick dashed red lines. Thin dashed red lines show $\pm 2\%$ deviation from $\chi_e^{ODT}$ in (a) and from $\chi_e^{crit}$ in (b). Snapshots of ordered and disordered states are included in (a) and (b). Panels (c) and (d) show normalized peak scattering intensities at each $T$ for the diBCP melt model (c) and the binary blend model (d).}
    \label{fig:4}
\end{figure*}

The $\chi_e(T)$ functions, computed at $\alpha_{ODT}(T=1)$ for the diBCP model and at $\alpha_{crit}(T=1)$ for the binary blend model, are shown in Fig. \ref{fig:4}a,b. For the diBCP melt model, $\chi_e(T)$ was strongly nonmonotonic and predicted ordering at $T\approx 1$, similar to the models in Fig. \ref{fig:3}, and further disordering at $T\approx 4.5$. This behavior was confirmed by the system snapshots and the $T$-dependency of the peak scattering intensity (Fig. \ref{fig:4}a,c). Therefore, the low-$\rho$ fully symmetric diBCP melt model exhibited closed-loop phase behavior. This behavior was predicted by the analytical theory in the $\rho\to 0$ limit (Fig. \ref{fig:1}b); however, the predicted onset of UCST point was predicted to happen at $T\gg a_{xx}/k_B=107$, which is much greater than $4.5$. As a result, we observed only qualitative agreement of the phase behavior of this model with the predictions of the $\rho\to 0$ theory.

On the other hand, the binary blend model exhibited no detectable phase transition (Fig. \ref{fig:4}b,d). This happened due to a smaller $\alpha_{crit}(T=1)$ compared to the value of $\alpha_{ODT}(T=1)$. In turn, this led to a much less pronounced $\chi_e(T)$ peak (Fig. \ref{fig:4}b), which lay within the $2\%$ margin from the phase separation point. This margin, in turn, corresponds to a typical width of the phase transition hysteresis within which the disordered state is metastable \cite{petrov2024simple}. As a result, Eq. \ref{eq1} predicted that the value of $\chi_e$ at the peak of $\chi_e(T)$ curve was insufficient to cause phase separation, and this prediction was confirmed by the system snapshots and the peak scattering intensity curve that was nearly flat within the error (Fig. \ref{fig:4}d).

\section{Discussion and Conclusions}

In this work, we discovered the appearance of LCST and closed-loop phase behavior in the simplest and widely used fully symmetric CG models of multicomponent liquids with arbitrary repulsive interaction potentials. This behavior was not expected from the previously developed theories; to clarify its origin, we derived the temperature dependence of the effective Flory-Huggins parameter $\chi_e$ in these models. We showed that if a model has a high monomer number density $\rho$, it can have a near-to-mean-field microscopic structure, which results in $\chi_e\propto 1/T$ and disordering upon heating. On the other hand, models with small $\rho$ have strong monomer-monomer correlations that make the mean-field description inaccurate, and the microscopic structure of the models is better approximated by an interacting dilute gas of CG monomers. In this regime, the theory predicted the \textit{growth} of $\chi_e$ with $T$ as $\chi_e\propto T^{1/m}$ in the $T\ll a_{xx}/k_B$ range, where $m>0$ is an integer, and the decrease with $T$ as $\chi_e\propto 1/T$ in the $T\gg a_{xx}/k_B$ range. As a result, the theory predicted the ordering upon heating and subsequent disordering after further increase of $T$, which meets the definition of the reentrant or "closed-loop" phase behavior. Importantly, the emergence of LCST/closed-loop phase behavior according to this mechanism did not depend on the exact form of repulsive nonbonded interaction potential and on compositional symmetry. As a result, these phenomena were predicted to occur even in the fully symmetric non-associating models.

To confirm these predictions, we constructed a model of fully symmetric binary blends and a model of fully symmetric diBCP melts with $\rho=5$ monomers per unit volume. In these models, the theoretical assumptions about the structure of high-$\rho$ models were accurate, which led to the observed UCST phase behavior (Fig. \ref{fig:2}). On the other hand, the models of the fully symmetric binary blends and diBCP melts with $\rho=0.7$ qualitatively satisfied the theoretical assumptions in the low-$\rho$ regime. This led to the LCST phase behavior in the short-bond models (Fig. \ref{fig:3}) and the closed-loop phase behavior in the models with longer bonds (Fig. \ref{fig:4}). Fig. \ref{fig:zinf_exponents} confirmed that for the harmonic nonbonded interaction potential used in DPD, $\rho\approx 1.0$ is the threshold below which the fully symmetric models have LCST and/or closed-loop phase behavior and not the UCST phase behavior expected from the previous theories.

The mechanism for the emergence of LCST and closed-loop phase behavior in this work is based on the rarification of monomer-monomer contacts in an arbitrary CG model of a multicomponent liquid. When a monomer has few (other-chain) neighbors with which it can interact, the radial distribution function obtains a strong temperature dependence: $g(r)\approx \exp[-U_\text{nonbonded}(r)/(k_BT)]$. As $T\to 0$, the probability of finding a neighboring monomer tends to zero inside the cutoff radius of the (repulsive) $U_\text{nonbonded}$, which leads to a superlinear $z_\infty(T)\to 0$ decrease. This superlinearity, in turn, means that the characteristic energy of excess repulsion between unlike particles ($z_\infty\alpha$) becomes smaller compared to the thermal energy $k_BT$ as $T$ decreases, which results in homogenization of the model liquid. Therefore, as $T\to 0$, $\chi_e\to 0$. Hence, in the low-$T$ limit, $\chi_e(T)$ is a monotonically increasing function, leading to the emergence of LCST. Due to the simplicity of the models in which this effect causes the non-UCST phase behavior, this mechanism might contribute to the temperature response of many experimental systems. Furthermore, it might be expected to play an important role in the systems that can be coarse-grained to a CG model with small density of "effective monomers" and a purely repulsive nonbonded interaction potential. It is worth mentioning that a low $\rho$ in a CG model does not imply unphysical gas-like conditions in the real system being modeled, since a CG "monomer" represents a large group of real atoms. Such systems may include diblock copolymer solutions or solutions of A/B homopolymers; however, determining the exact range of these systems requires systematic coarse-graining that also preserves the temperature response of the original system in the CG model, which is outside of the scope of this paper and a interesting topic for future research. Finally, it is worth noting that the mechanism for LCST/closed-loop phase behavior described above is relevant in the studied type of CG models and does not generalize other previously described mechanisms that are relevant in other models (i.e., with certain types of association or the asymmetry of equation of state properties between species). The possible existence of its "effective" equivalence to other mechanisms is a good topic for future research.

\section*{Acknowledgements}

We acknowledge the MIT SuperCloud and Lincoln Laboratory Supercomputing Center for providing high performance computing resources. This research was supported by the National Science Foundation through award number DMREF 2118678.

\setcounter{page}{1}

\section*{Supplemental Material}
\addcontentsline{toc}{section}{Supplemental Material}
\clearpage

\setcounter{figure}{0}
\setcounter{table}{0}
\setcounter{equation}{0}
\setcounter{section}{0}

\renewcommand{\thefigure}{S\arabic{figure}}
\renewcommand{\thetable}{S\arabic{table}}
\renewcommand{\theequation}{S\arabic{equation}}
\renewcommand{\thesection}{S\arabic{section}}

\input{SIarxiv}

\clearpage
\bibliographystyle{unsrt}
\bibliography{sample}
\end{document}

%% file: SIarxiv.tex
 \begin{figure}[!t]
\centering
  \begin{subfigure}{0.325\textwidth}
	\includegraphics[width=\linewidth,height=\textheight,keepaspectratio]{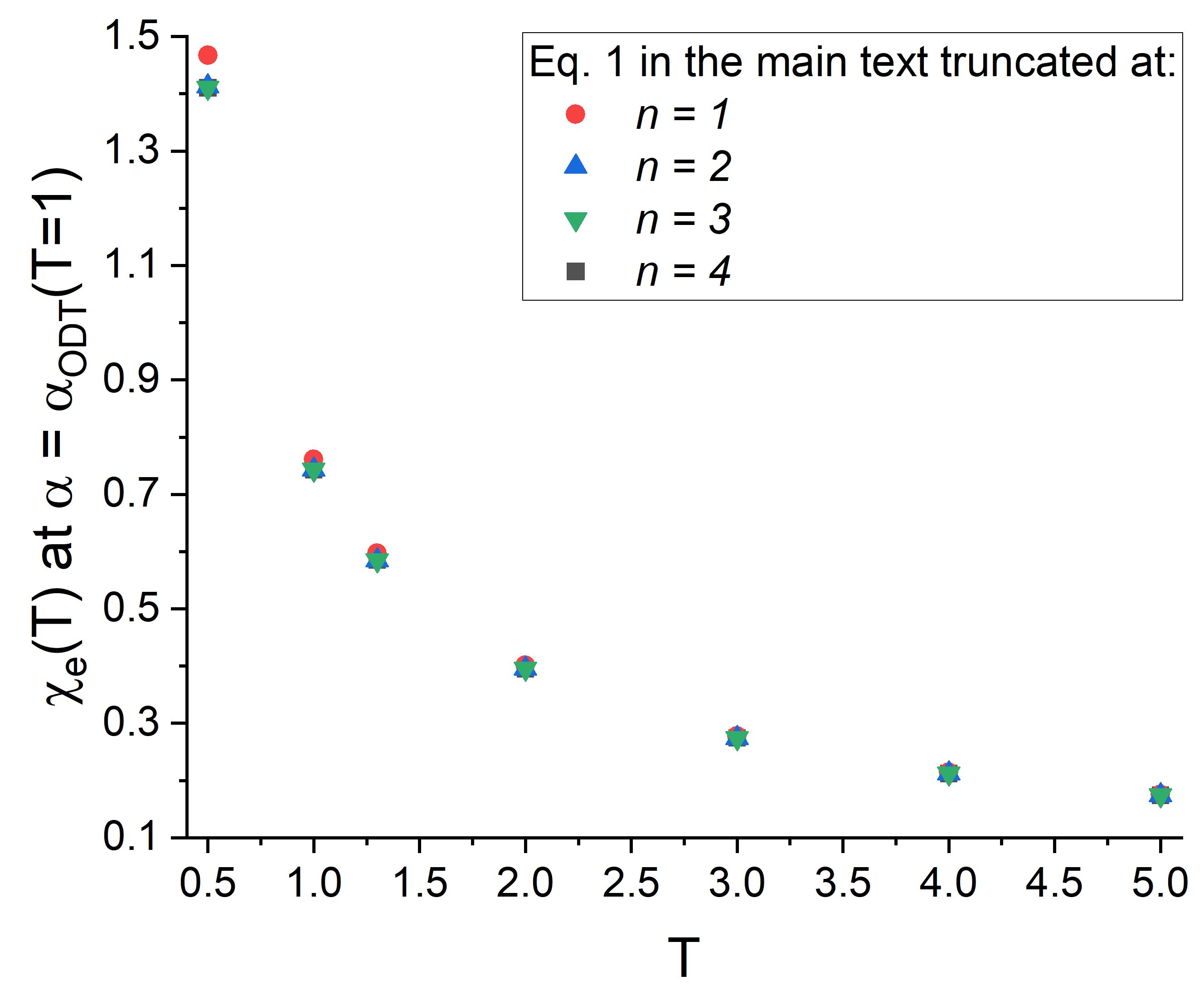}
	\caption{}
	\end{subfigure}
	\begin{subfigure}{0.325\textwidth}
	\includegraphics[width=\linewidth,height=\textheight,keepaspectratio]{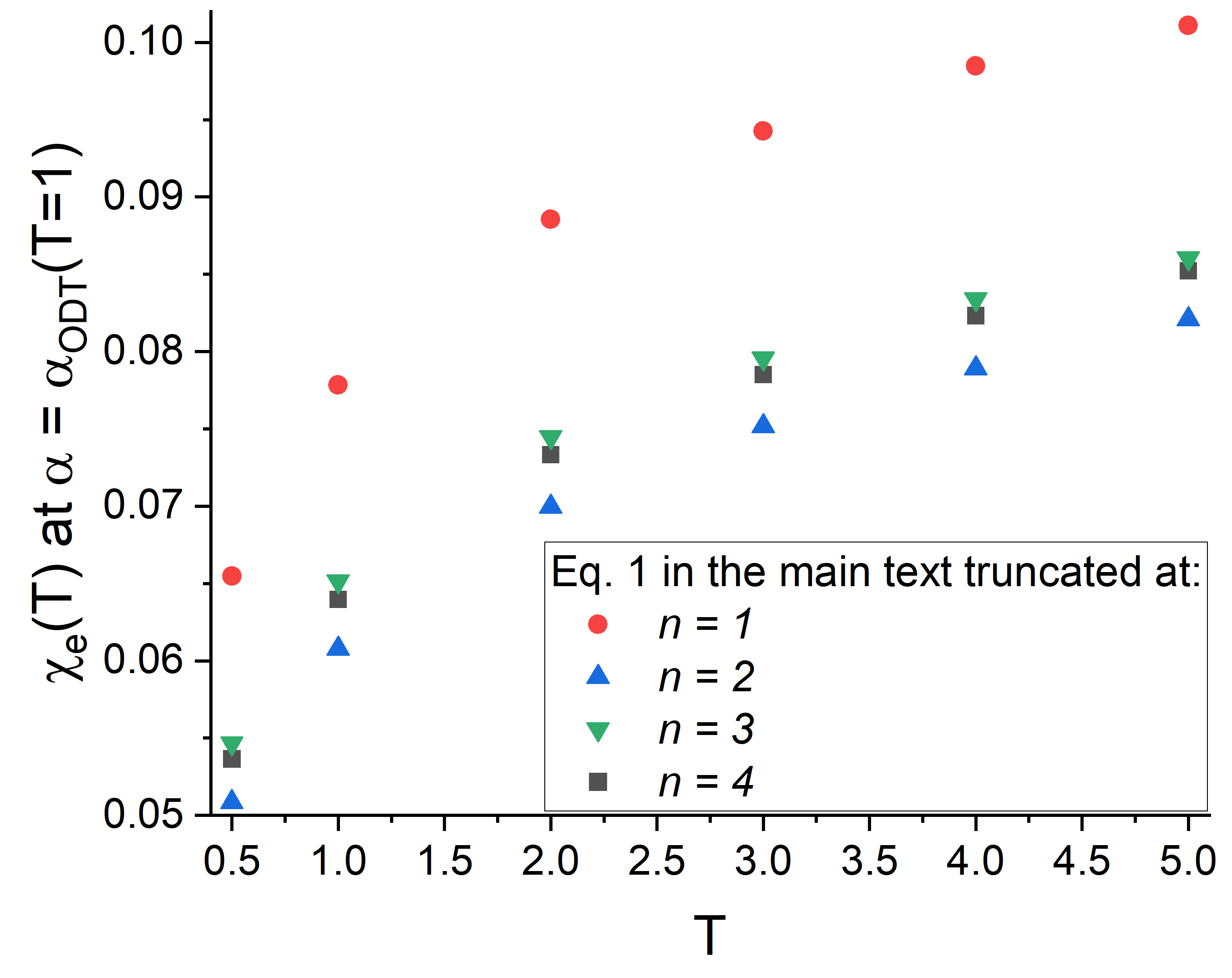}
	\caption{}
	\end{subfigure}
    \begin{subfigure}{0.325\textwidth}
	\includegraphics[width=\linewidth,height=\textheight,keepaspectratio]{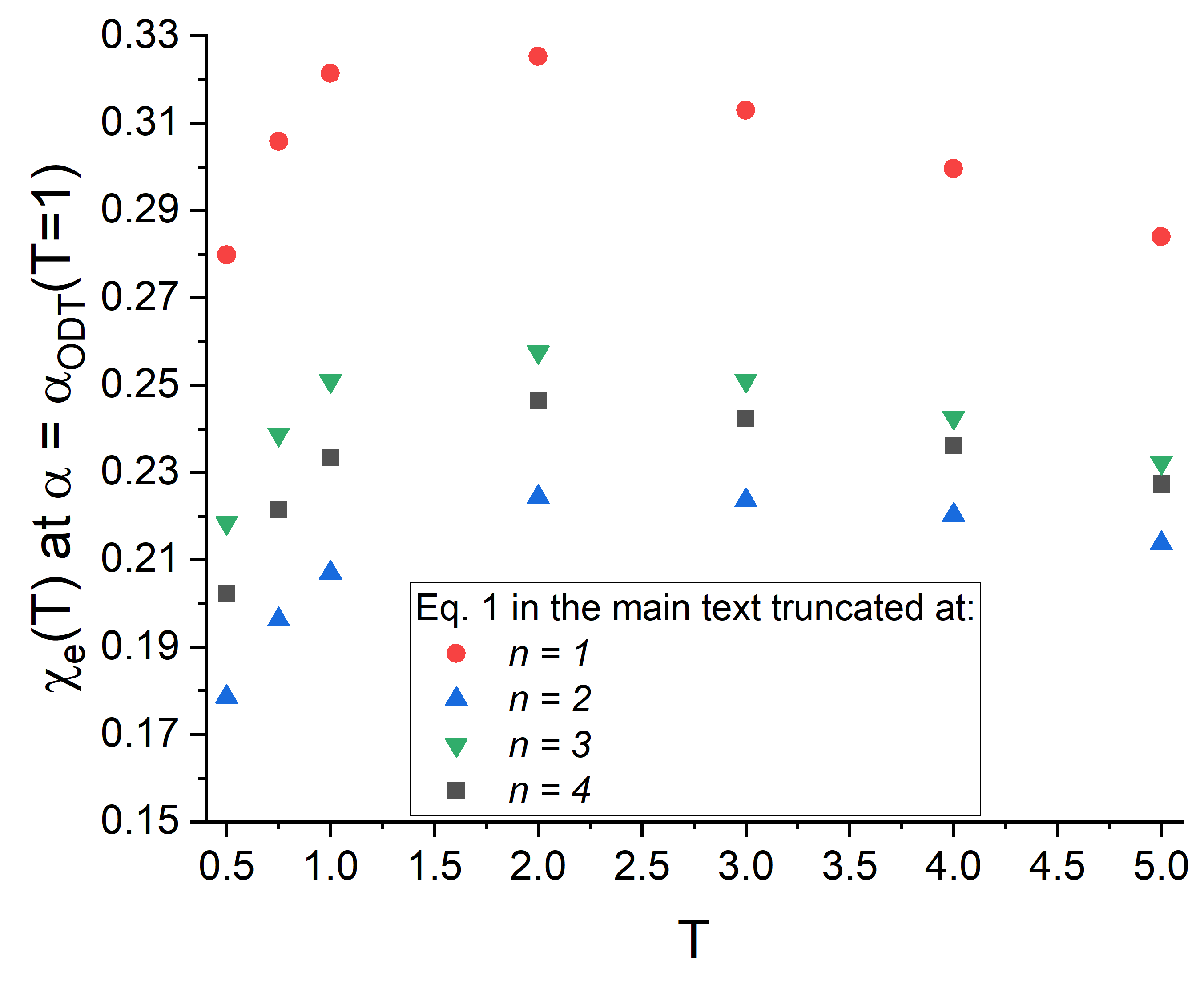}
	\caption{}
	\end{subfigure}
  \caption{$T$-dependence of $\chi_e$ estimated using different truncations (at $n=1$ (red), $n=2$ (blue), $n=3$ (green), and $n=4$ (grey)) of Eq. 1 in the main text. (a) high-$\rho$ ($\rho=5$) model, (b) low-$\rho$ model ($\rho=0.7$) with $r_0=1.0$, and (c) low-$\rho$ model ($\rho=0.7$) with $r_0=1.5$. For each model, $\alpha$ value was set to the maximal value of $\alpha$ used in this work, i.e., to $\alpha_{ODT}$ at $T=1$ (see main text). This ensured maximal possible discrepancy between the true value of $\chi_e$ (grey points) and the linear $\chi_e=z_\infty\alpha/k_BT$ approximation used to estimate the $\chi_e(T)$ function analytically in section 2 of the main text (red points).}
\label{fig_si:approximations}
\end{figure}

 \begin{figure}[!t]
\centering
  \begin{subfigure}{0.325\textwidth}
	\includegraphics[width=\linewidth,height=\textheight,keepaspectratio]{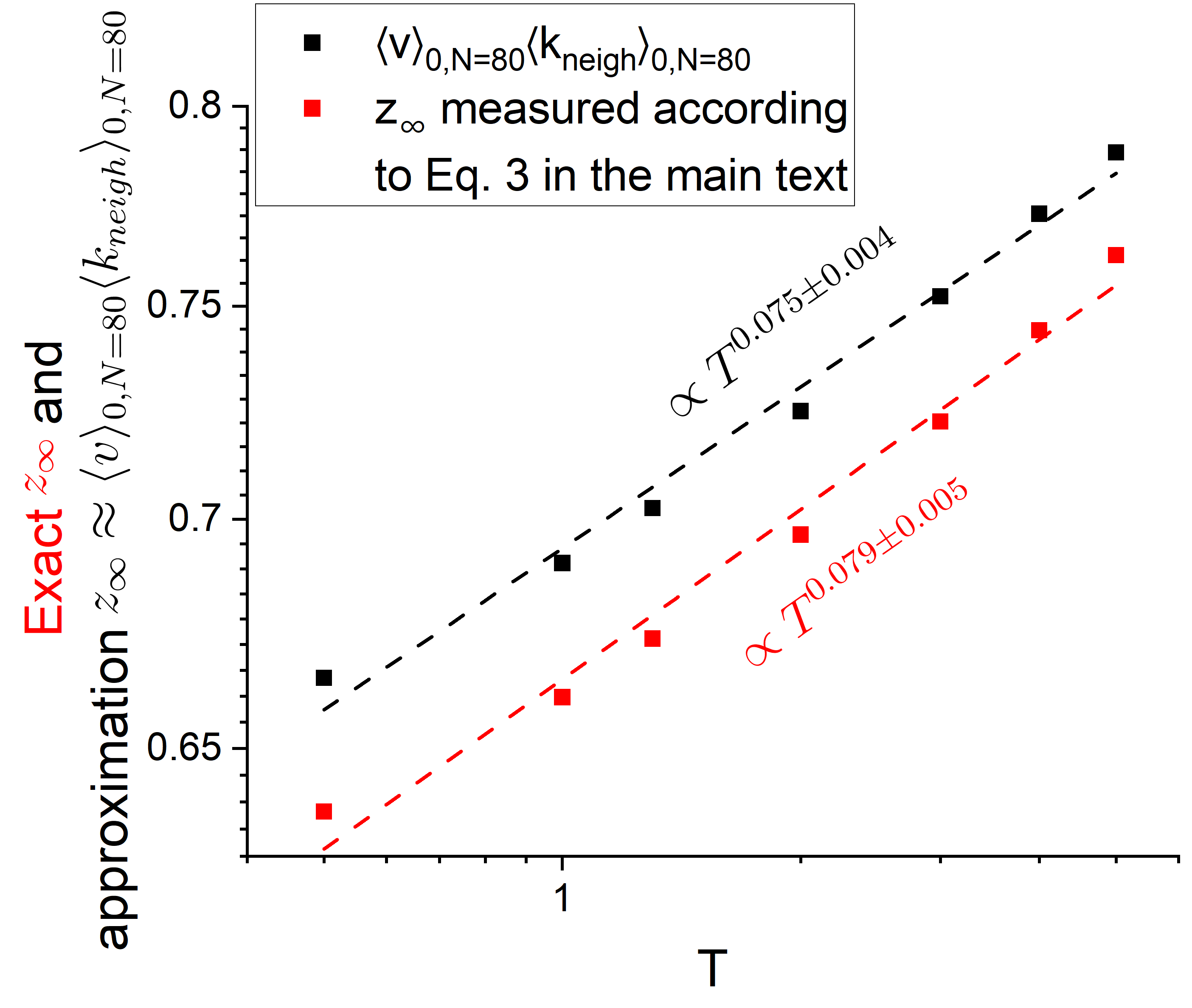}
	\caption{}
	\end{subfigure}
	\begin{subfigure}{0.325\textwidth}
	\includegraphics[width=\linewidth,height=\textheight,keepaspectratio]{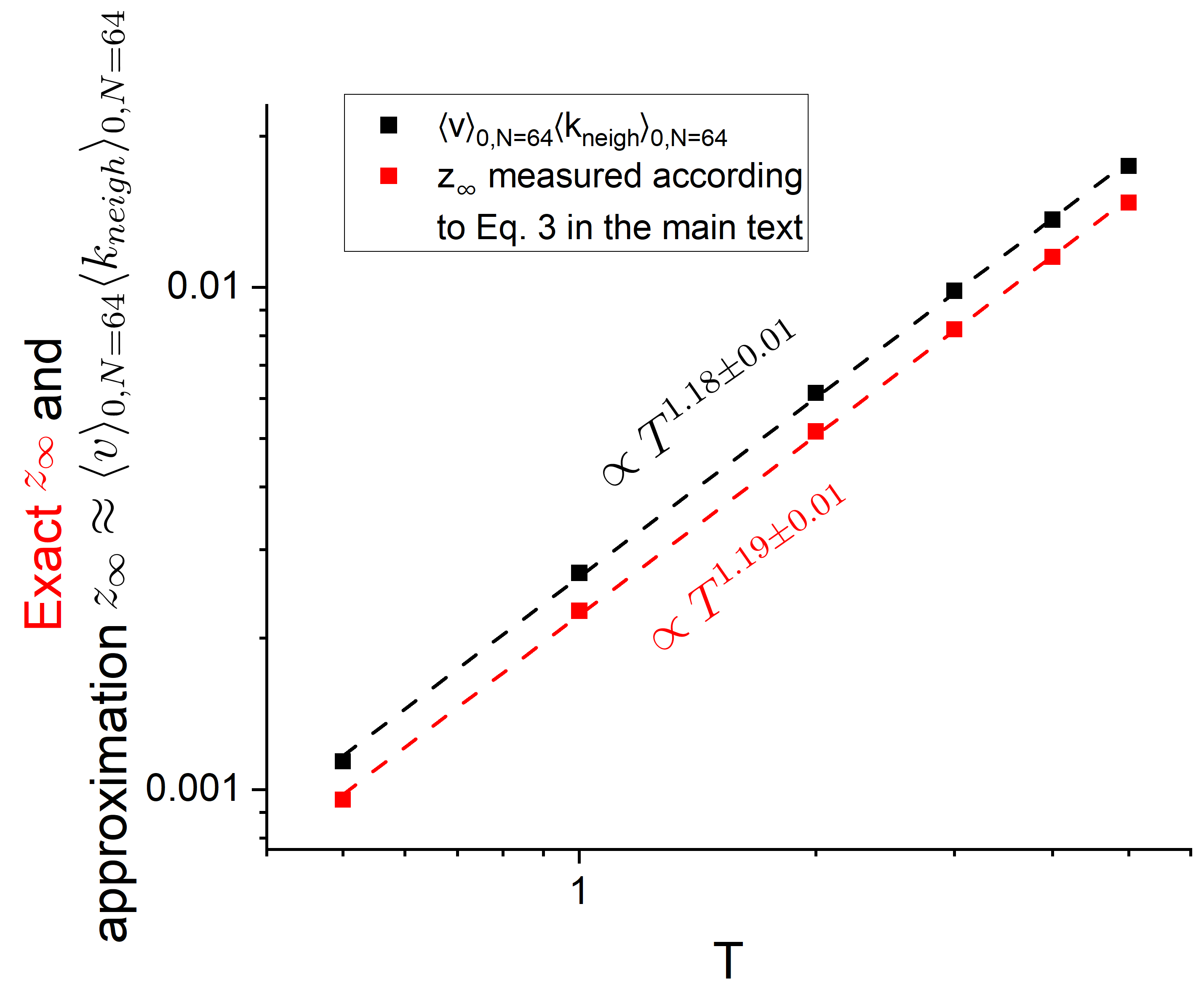}
	\caption{}
	\end{subfigure}
    \begin{subfigure}{0.325\textwidth}
	\includegraphics[width=\linewidth,height=\textheight,keepaspectratio]{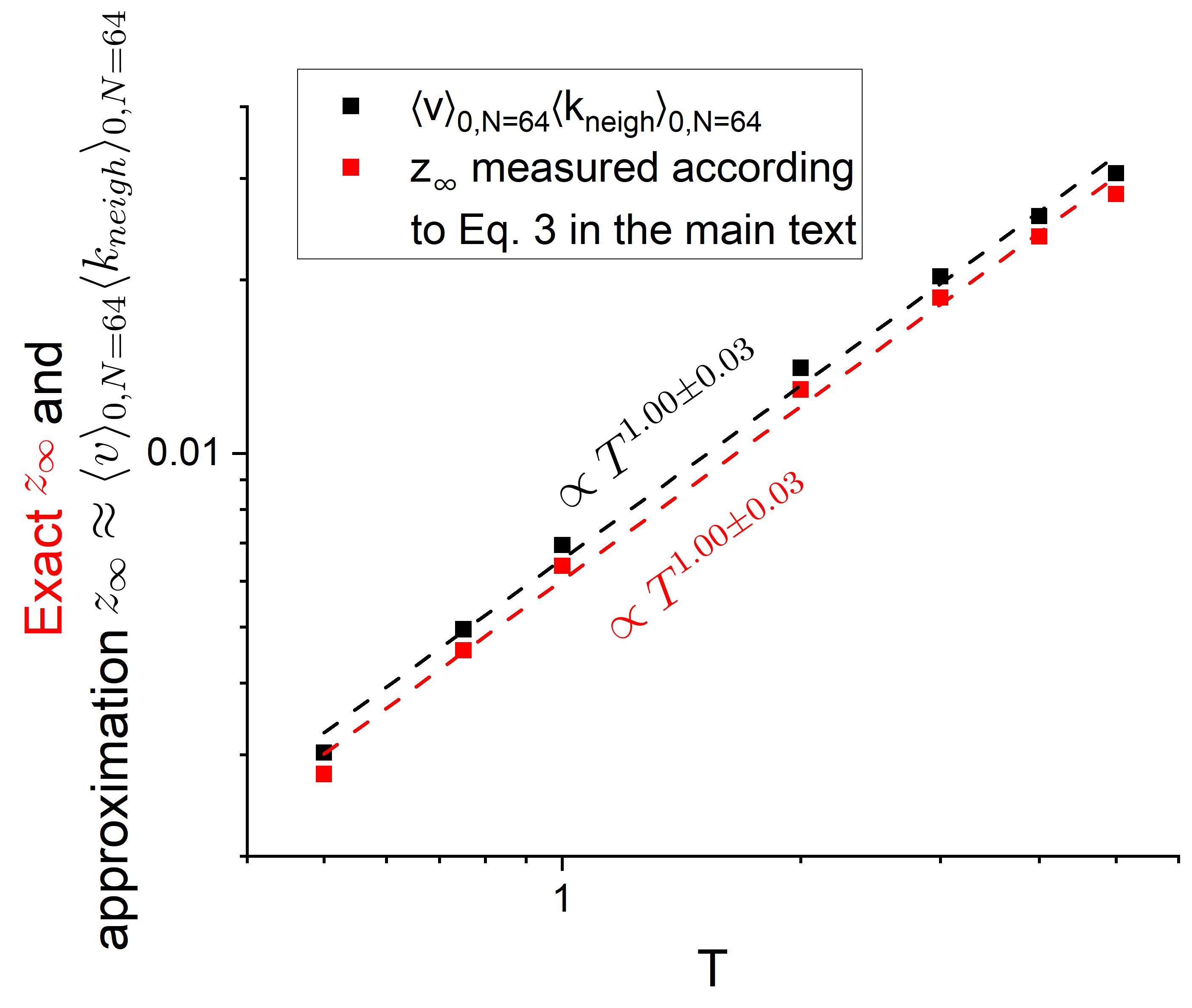}
	\caption{}
	\end{subfigure}
  \caption{$T$-dependencies of the exact value of $z_\infty$ calculated according to Eq. 3 in the main text (red) and of the approximation for $z_\infty$ used in the analytical theory (section 2 in the main text) $z_\infty\approx \langle v\rangle\langle k_{neigh}\rangle$ (black). For black points, averaging was performed at $\alpha=0$ and at largest simulated $N$ to mimic the required reference state conditions ($\alpha=0$, $N\to\infty$). (a) high-$\rho$ ($\rho=5$) model, (b) low-$\rho$ model ($\rho=0.7$) with $r_0=1.0$, and (c) low-$\rho$ model ($\rho=0.7$) with $r_0=1.5$.}
\label{fig_si:zinf_factorization}
\end{figure}

\begin{figure}[!t]
\centering
  \begin{subfigure}{0.49\textwidth}
	\includegraphics[width=\linewidth,height=\textheight,keepaspectratio]{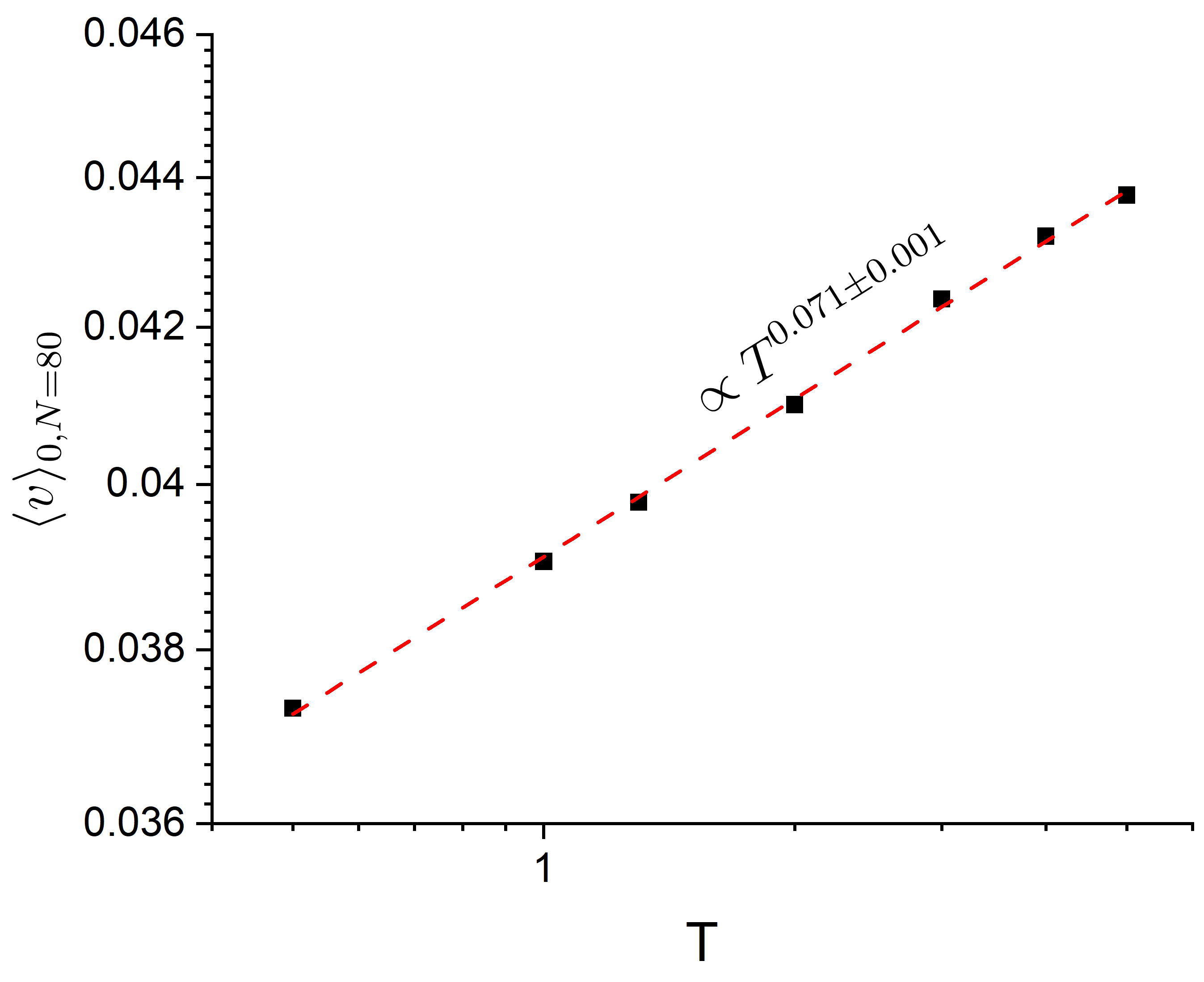}
	\caption{}
	\end{subfigure}
	\begin{subfigure}{0.49\textwidth}
	\includegraphics[width=\linewidth,height=\textheight,keepaspectratio]{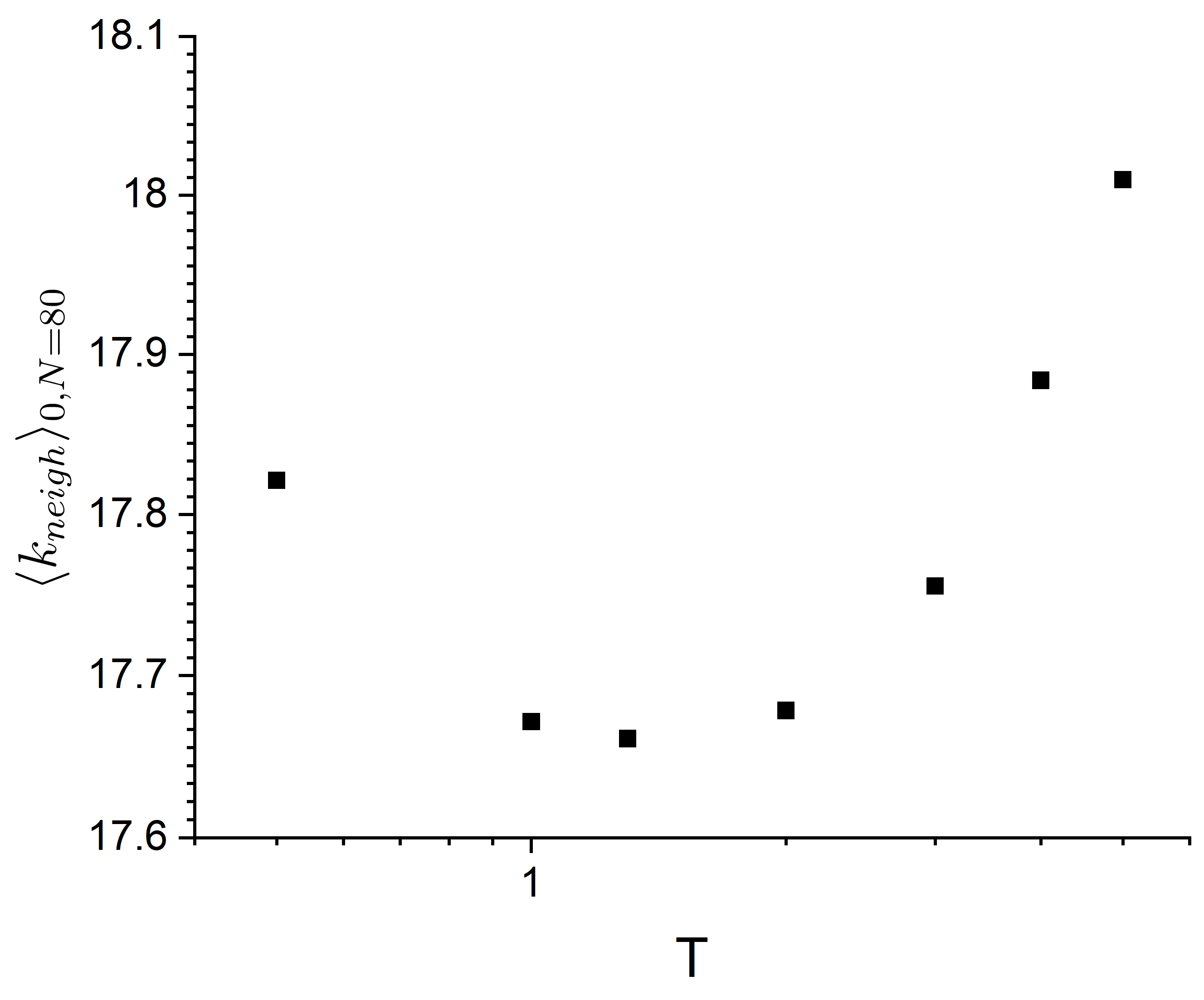}
	\caption{}
	\end{subfigure}
  \caption{$T$-dependence of (a) the average value of $v(r)$ and (b) the number of other-chain neighbors $k_{neigh}$ for the high-$\rho$ ($\rho=5$) fully symmetric models of diBCP melts and binary blends. Averaging was performed at $\alpha=0$ and at largest simulated $N$ to mimic the required reference state conditions ($\alpha=0$, $N\to\infty$). In (a), red dashed line denotes the best power law fit; in (b), no such fit was performed due to the absence of a clear $\langle k_{neigh}\rangle_{0,N=80}(T)$ trend.}
\label{fig_si:vk_ucst}
\end{figure}

\begin{figure}[!t]
\centering
  \begin{subfigure}{0.49\textwidth}
	\includegraphics[width=\linewidth,height=\textheight,keepaspectratio]{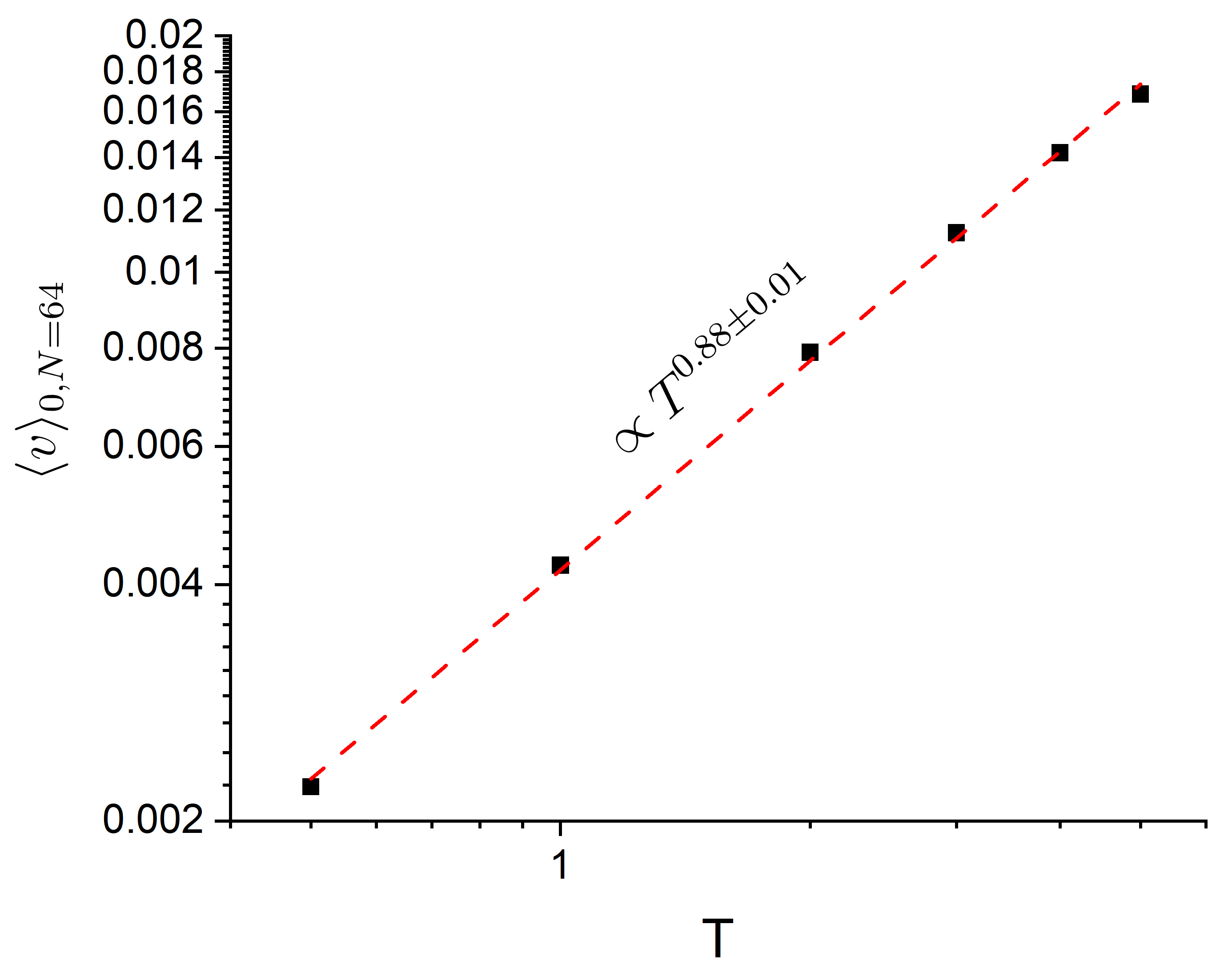}
	\caption{}
	\end{subfigure}
	\begin{subfigure}{0.49\textwidth}
	\includegraphics[width=\linewidth,height=\textheight,keepaspectratio]{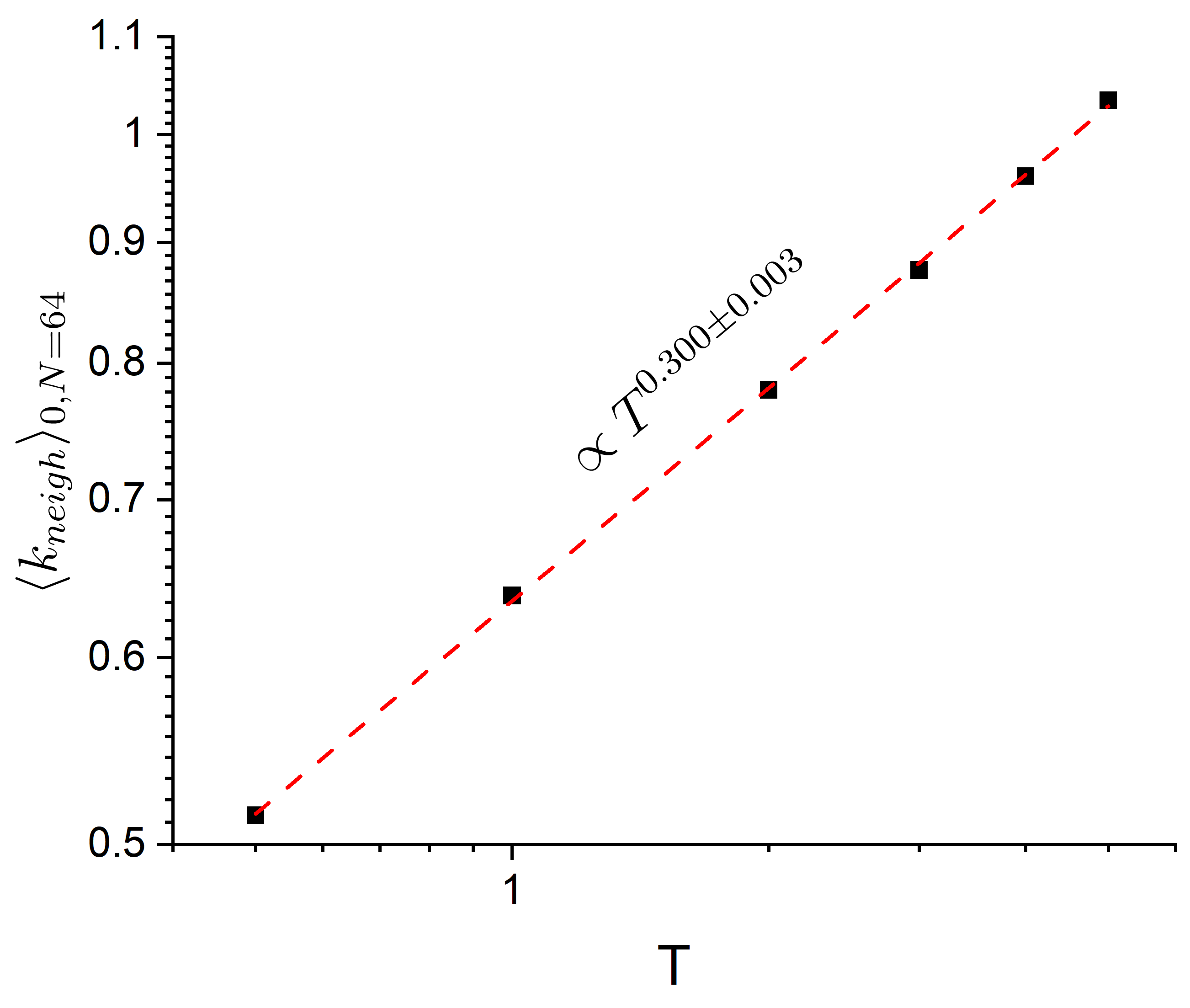}
	\caption{}
	\end{subfigure}
  \caption{$T$-dependence of (a) the average value of $v(r)$ and (b) the number of other-chain neighbors $k_{neigh}$ for the low-$\rho$ ($\rho=0.7$, $r_0=1.0$) fully symmetric models of diBCP melts and binary blends. Averaging was performed at $\alpha=0$ and at largest simulated $N$ to mimic the required reference state conditions ($\alpha=0$, $N\to\infty$). Red dashed lines denote the best power law fit.}
\label{fig_si:vk_lcst}
\end{figure}

\begin{figure}[!t]
\centering
  \begin{subfigure}{0.49\textwidth}
	\includegraphics[width=\linewidth,height=\textheight,keepaspectratio]{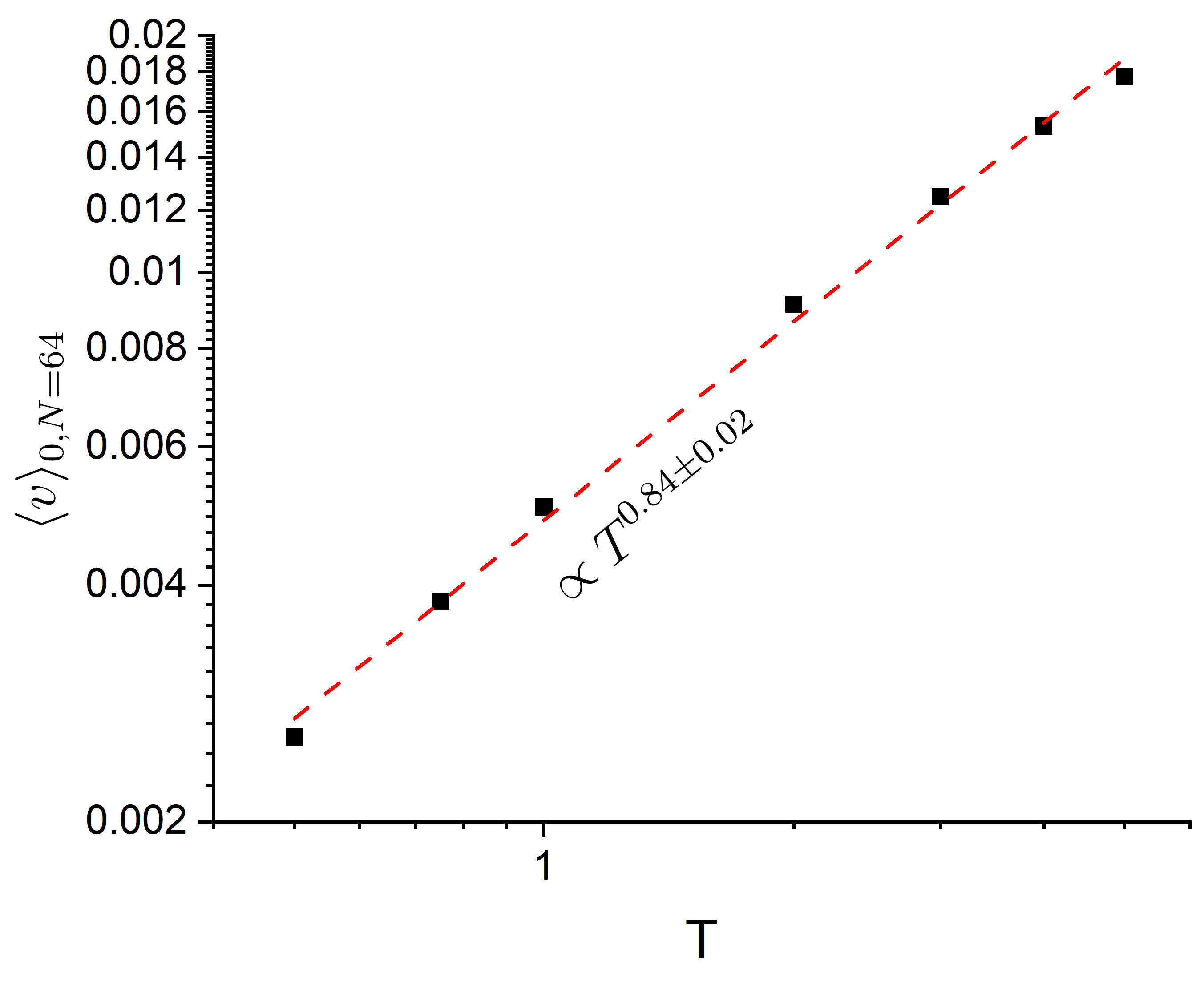}
	\caption{}
	\end{subfigure}
	\begin{subfigure}{0.49\textwidth}
	\includegraphics[width=\linewidth,height=\textheight,keepaspectratio]{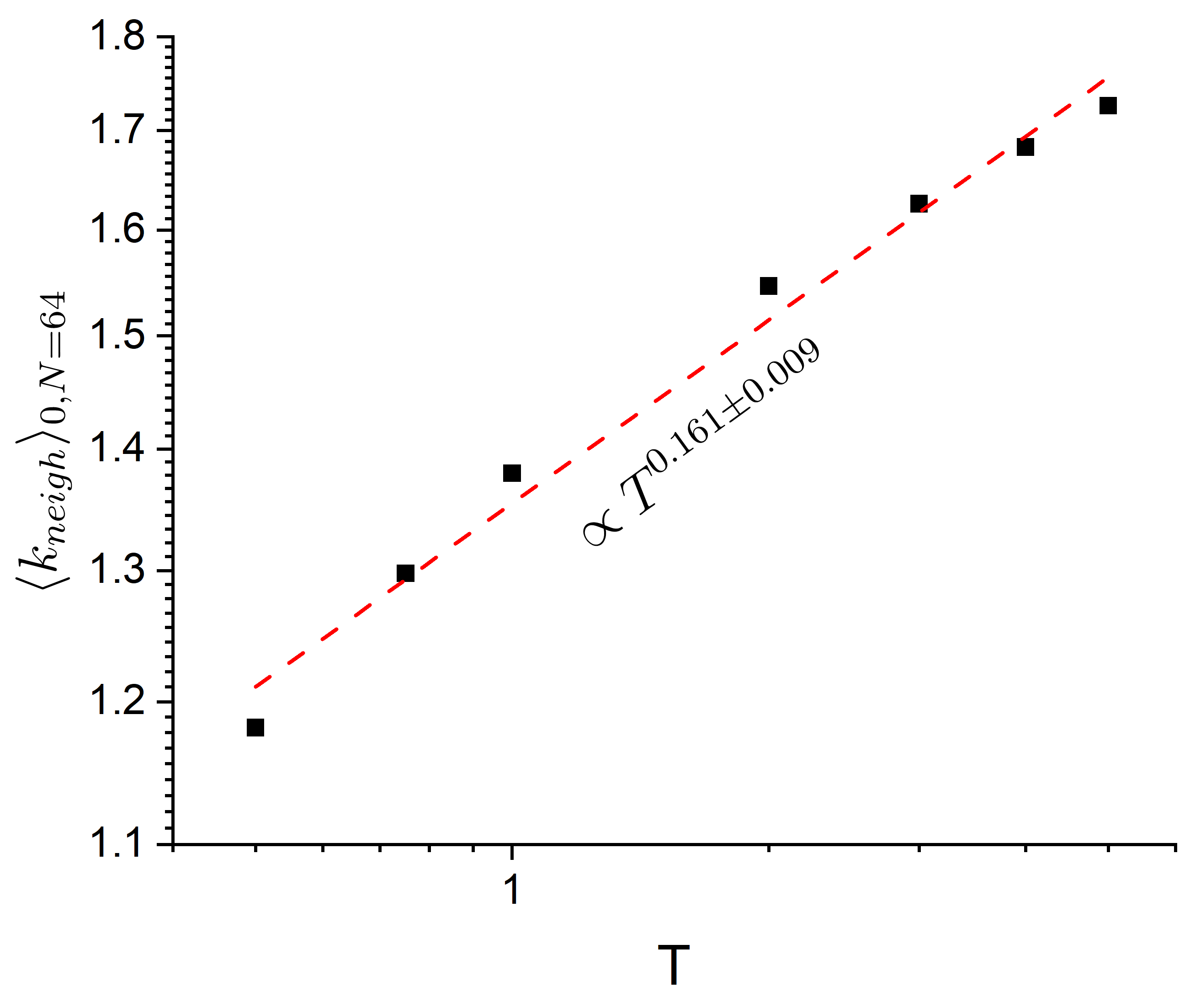}
	\caption{}
	\end{subfigure}
  \caption{$T$-dependence of (a) the average value of $v(r)$ and (b) the number of other-chain neighbors $k_{neigh}$ for the low-$\rho$ ($\rho=0.7$, $r_0=1.5$) fully symmetric models of diBCP melts and binary blends. Averaging was performed at $\alpha=0$ and at largest simulated $N$ to mimic the required reference state conditions ($\alpha=0$, $N\to\infty$). Red dashed lines denote the best power law fit.}
\label{fig_si:vk_closedloop}
\end{figure}